\documentclass[pdflatex,referee,sn-basic]{sn-jnl}% referee option is meant for double line spacing

\newcommand*\patchAmsMathEnvironmentForLineno[1]{%
  \expandafter\let\csname old#1\expandafter\endcsname\csname #1\endcsname
  \expandafter\let\csname oldend#1\expandafter\endcsname\csname end#1\endcsname
  \renewenvironment{#1}%
     {\linenomath\csname old#1\endcsname}%
     {\csname oldend#1\endcsname\endlinenomath}}%
\newcommand*\patchBothAmsMathEnvironmentsForLineno[1]{%
  \patchAmsMathEnvironmentForLineno{#1}%
  \patchAmsMathEnvironmentForLineno{#1*}}%
\AtBeginDocument{%
\patchBothAmsMathEnvironmentsForLineno{equation}%
\patchBothAmsMathEnvironmentsForLineno{align}%
\patchBothAmsMathEnvironmentsForLineno{flalign}%
\patchBothAmsMathEnvironmentsForLineno{alignat}%
\patchBothAmsMathEnvironmentsForLineno{gather}%
\patchBothAmsMathEnvironmentsForLineno{multline}%
}
 \usepackage[switch]{lineno} % For 2-column

\jyear{2023}%

\theoremstyle{thmstyleone}%
\theoremstyle{thmstyletwo}%
\theoremstyle{thmstylethree}%

\newif\ifUseOverleaf
 \UseOverleaftrue

\ifUseOverleaf
\else
  \usepackage{mediabb}
\fi
\usepackage{bm}
\usepackage{ulem}
\DeclareRobustCommand{\erase}{\bgroup\markoverwith{\textcolor{red}{\rule[.5ex]{2pt}{1.2pt}}}\ULon}

\def\myfigwidthF{0.960\textwidth}  % for the 1-column style
\def\myfigwidthf{0.880\textwidth}  % for the 1-column style, narrower
\long\def\comment#1{}
\comment{
   Usage:
     \comment{ comments....}
}

\def\app#1#2{%
  \mathrel{%
    \setbox0=\hbox{$#1\sim$}%
    \setbox2=\hbox{%
      \rlap{\hbox{$#1\propto$}}%
      \lower1.1\ht0\box0%
    }%
    \raise0.25\ht2\box2%
  }%
}

\usepackage{marginnote}
 at 7truept

 \definecolor{applegreen}{rgb}{0.55,0.71,0.0}
 \definecolor{forestgreen}{rgb}{0.13,0.55,0.13}
 \definecolor{pinegreen}{rgb}{0.0,0.47,0.44}
 \definecolor{upforestgreen}{rgb}{0.0,0.27,0.13}
 \definecolor{vividviolet}{rgb}{0.62,0.0,1.0}
 \definecolor{lightcyan}{rgb}{0.717647, 0.933071, 0.996078} % in RGB
 \definecolor{scarlet}{rgb}{0.7421875, 0.00390625, 0.09765625}

 \def\revisedtextcolor{black}

 \newcommand{\revisedtext}[1]{{\leavevmode\textcolor{\revisedtextcolor}{#1}}}

\begin{document}

\title[Pluto's argument of perihelion and the major planets]{Libration of Pluto's argument of perihelion and the role of the major planets}

\author*[1,2,3]{\fnm{Takashi} \sur{Ito}}\email{tito@cfca.nao.ac.jp}
\author[4]{\fnm{Renu} \sur{Malhotra}}\email{malhotra@arizona.edu}

\affil*[1]{\orgdiv{Center for Computational Astrophysics}, \orgname{National Astronomical Observatory}, \orgaddress{\street{Osawa 2--21--1}, \city{Mitaka}, \postcode{181--8588}, \state{Tokyo}, \country{Japan}}}
\affil[2]{\orgdiv{Planetary Exploration Research Center}, \orgname{Chiba Institute of Technology}, \orgaddress{\street{2--17--1}, \city{Narashino}, \postcode{275--0016}, \state{Chiba}, \country{Japan}}}
\affil[3]{\orgdiv{College of Science and Engineering}, \orgname{Chubu University}, \orgaddress{\street{1200 Matsumoto--cho}, \city{Kasugai}, \postcode{487--8501}, \state{Aichi}, \country{Japan}}}
\affil[4]{\orgdiv{Lunar and Planetary Laboratory}, \orgname{The University of Arizona}, \orgaddress{\street{1629 E University Blvd}, \city{Tucson}, \postcode{85721}, \state{AZ}, \country{USA}}}

\abstract{%
Pluto's argument of perihelion is known to librate around $90^\circ$. 
This libration is related to the secular phenomenon known as the von Zeipel--Lidov--Kozai (vZLK) oscillation.
In this work, we make a quantitative assessment of the influence of Neptune's mean motion resonance and of the other giant planets' secular perturbations on the libration of Pluto's argument of perihelion.
Here, a parameter $k^2 = (1-e^2) \cos^2 I$ is the key where $e$ is eccentricity and $I$ is inclination.
When $k^2$ of a Pluto-like object is larger than a certain critical value, libration of its argument of perihelion would not occur.
The secular effect of other disturbing planets (Jupiter, Saturn, Uranus) plays a significant role in determining the critical $k^2$.
The non-zero oscillation amplitude of the critical resonant argument also plays a role, although not a dominant one.

}% End of \abstract


\maketitle

\section{Introduction}\label{sec:intro}
Pluto's orbit is peculiar.
It is known to be stable, and its long-term stability involves two major dynamical characteristics.
One of them is the 3:2 mean motion resonance with Neptune.
When a three-body system is in a mean motion resonance, the mean longitudes of the objects are not independent, and are linked through the librations of the critical resonant argument $\sigma$ defined as follows, % generally has a finite amplitude of oscillation.
\begin{equation}
  \sigma = p \lambda - p' \lambda' - (p-p') \varpi ,
  \label{eqn:def-sigma}
\end{equation}
with integers $p > p' > 0$
where $\lambda$ and $\varpi$ denote perturbed body's mean longitude and its longitude of perihelion, and $\lambda'$ denotes perturbing body's mean longitude.
For Pluto--Neptune's 3:2 exterior mean motion resonance, we have $p=3$ and $p'=2$, therefore $ \sigma = 3 \lambda - 2 \lambda' - \varpi $.
$\sigma$ in the Pluto--Neptune system librates around $180^\circ$ with a period of about 20,000 years and with an amplitude of approximately $85^\circ$ to $90^\circ$ \citep[e.g.][]{cohen1965,malhotra1997}.
The other characteristics of Pluto's motion is the libration of its argument of perihelion, $g$, around $90^\circ$ with an amplitude of about $25^\circ$ with a period of about $4$ million years \citep[e.g.][]{kinoshita1996a}.
The $g$-libration, which is an aspect of the so-called von Zeipel--Lidov--Kozai oscillation \citep[][hereafter abbreviated as vZLK oscillation]{vonzeipel1910,lidov1961-en,kozai1962b,shevchenko2017,ito2019}, ensures that Pluto's location is well above the plane of the other planets when Pluto approaches its perihelion.
These two characteristics keep Pluto's minimum distance from Neptune relatively large, promoting its very long term stability \citep[e.g.][]{ito2002a,zaveri2021}.
Note that throughout the present study, we use $g$ rather than $\omega$ for denoting argument of pericenter or perihelion.
$g$ is one of the Delaunay elements, and we use it following the pioneering work along this line of studies \citep[e.g.][]{vonzeipel1910}.

The theoretical framework for the vZLK oscillation is the doubly averaged circular restricted three-body problem (hereafter referred to as CR3BP).
The semimajor axis of the perturbed object becomes constant after the averaging procedure.
The vertical ($z$-) component of the angular momentum of the perturbed object also becomes constant in the doubly averaged CR3BP
if the $z$-axis of the coordinates is aligned with the orbit pole of the perturbing object.
Therefore we can define the following dimensionless parameter which we use throughout this study:
\begin{equation}
  k^2 = \left( 1-e^2 \right) \cos^2 I ,
  \label{eqn:def-k2}
\end{equation}
where $e$ is eccentricity of the perturbed body, and $I$ is its inclination to the orbital plane of the perturbing object.
For bound orbits, the full range of $k^2$ is $(0,1)$.
Here we follow the notation in \citet{vonzeipel1910} and use $k^2$ for this parameter.

In the doubly averaged CR3BP, it is known that there is a critical value of $k^2$ above which the argument of pericenter $g$ of the perturbed body cannot librate.
Hereafter we call it $k^2_\mathrm{crit}$ in this study.
The value of $k^2_\mathrm{crit}$ depends on the ratio of semimajor axis $(a)$ of the perturbed body to that of the perturbing body $(a')$.
In the inner CR3BP in which the perturbed body's orbit is located interior to the perturbing body's orbit, $k^2_\mathrm{crit} \to 0.6 \; (= \frac{3}{5})$ in the limit of $\alpha \equiv a/a' \to 0$.
In the outer CR3BP where the perturbed body's orbit is located exterior to the perturbing body's orbit,
$k^2_\mathrm{crit} \to 0.2\; (= \frac{1}{5})$ in the limit of $\alpha' \equiv a'/a \to 0$.
The full range of the $\alpha$ or $\alpha'$ is $(0,1)$.
In more general cases where $\alpha$ or $\alpha'$ is not negligibly small, the larger $\alpha$ or $\alpha'$ is, the greater $k^2_\mathrm{crit}$ becomes.

Pluto's orbit has the parameter value of $k^2 \simeq 0.871$.
This is obtained with the use of $e=0.249018$ and $I=15^\circ.5639$ which are the osculating eccentricity and inclination of the Pluto system barycenter with respect to the solar system barycenter (obtained from the Horizons System of JPL on April 12, 2024), and $I$ is measured from the solar system's invariable plane \citep{souami2012}.
We make this choice of reference plane because the averaged Hamiltonian model employed in this study assumes that all the major planets move on the same plane.
This is equivalent to the solar system invariable plane by definition.
Following \citet{malhotra2025}, we can also estimate Pluto's value of $k^2$ from its time-averaged orbital elements (averaged from the numerical integration of 100 myr):
$\left< e \right> \simeq  0.24373$,
$\left< I \right> \simeq 15.937^\circ$;
then we find $k^2 \simeq 0.870$, very close to the value obtained from the osculating orbital elements.

Pluto with Neptune as the perturbing planet composes the outer CR3BP having $\alpha' \simeq 0.763$.
In the doubly averaged (non-resonant) CR3BP, $k^2_\mathrm{crit} \approx 0.30$ for $\alpha' \sim 0.763$ \citep{ito2019}.
Since Pluto's $k^2 \approx 0.87$ as mentioned above, this is quite insufficient to fulfill the condition $k^2 < k^2_\mathrm{crit}$ for Pluto's $g$ to librate; there must be some other dynamical effects.
There are two of them that we have to consider.
First, Pluto is trapped in the 3:2 mean motion resonance with Neptune.
Mean motion resonances in general could modify the structure of the phase space and change the critical values of $k^2$.
Second,
previous studies with classical perturbation method as well as numerical propagation already showed that the secular effect of the other planets (Jupiter, Saturn and Uranus) is significant for Pluto’s long term orbit evolution \citep[e.g.][]{hori1968,nacozy1978a,kinoshita1996a}.
Our recent numerical explorations identified in more detail the specific role of the other giant planets for supporting the conditions for Pluto’s $g$ libration and in protecting Pluto from chaos in its phase space neighborhood \citep{malhotra2022}. 
In the present study, we extend that analysis by employing numerical quadrature to further elucidate the dependence of Pluto’s $g$-libration on the parameters of the other giant planets as well as on the amplitude of the critical resonant argument.

In Section \ref{sec:quadrature}, we discuss how we employ numerical quadrature for estimating $k^2_\mathrm{crit}$ of perturbed object in the averaged CR3BP while including the effect of the 3:2 mean motion resonance.
In Section \ref{sec:otherplanets}, we introduce the oblate Sun model to incorporate the influence of the major disturbing planets other than Neptune.
Section \ref{sec:discussion} provides a summary and discussion.

\section{Numerical quadrature}\label{sec:quadrature}
\subsection{Purpose of the procedure\label{ssec:quadrature-purpose}}
In this section we describe how we estimate $k^2_\mathrm{crit}$ in the averaged CR3BP.
Our principle is simple: we evaluate the values of the disturbing function (hereafter denoted as $R$) by numerical quadrature, and draw its averaged values on a two dimensional phase space.
$R$ is defined in the equation of motion of the three-body system
\begin{equation}
  \frac{d^2 \bm{r}}{dt^2} + \mu \frac{\bm{r}}{r^3} = \nabla R ,
  \label{eqn:EOM-relative-inner-pre}
\end{equation}
with
\begin{equation}
  R = \frac{\mu'}{\left\lvert \bm{r}' - \bm{r} \right\rvert}
     - \mu' \frac{\bm{r}' \cdot \bm{r}}{r'^3},
  \label{eqn:def-R}
\end{equation}
where
$\bm{r}$ indicates the position vector of the perturbed body,
$\bm{r}'$ is the position vector of the perturbing body,
$\mu' = {\cal G} m'$,
$m'$ is perturbing body's mass,
${\cal G}$ is the gravitational constant,
$\mu = {\cal G} (m_0 + m)$,
$m_0$ is the central mass, and
$m$ is perturbed body's mass which we assume is negligibly small.
The value of $R$ is evaluated using the osculating orbital elements of the perturbed object.
Then, we carry out numerical averaging of $R$ over fast-rotating variables.
The averaged $R$ is a function of $g$ and $e$ in CR3BP, and can be denoted as $\overline{R}(g,e)$.
One of the reasons for this is that the semimajor axis $a$ is constant in the averaged system.
This comes from the fact
that when a certain angular variable (e.g. mean anomaly) is averaged and removed in a canonical transformation employed in perturbation method, the corresponding conjugate momentum (e.g. a Delaunay element $L$ which is a function of $a$) will remain constant.
The second reason is that inclination $I$ is connected to $e$ through the relationship Eq. \eqref{eqn:def-k2}.
Consequently, $a$ and $k^2$ are implicitly the free parameters in $\overline{R}(g,e)$.
$\overline{R}$ does not depend on longitude of ascending node of the perturbed body either, a consequence of the azimuthal symmetry of the disturbing potential when averaged over the perturbing body's orbit.

Here is the specific way how we estimate $k^2_\mathrm{crit}$ in this study.
Suppose a local extremum of $\overline{R}(g,e)$ shows up somewhere on the phase space when $k^2 = k_1^2$.
Then, if the local extremum vanishes when we slightly increase the value of $k^2$ from $k_1^2$ to $k_2^2$, we can say that $k^2_\mathrm{crit}$ of this system at the given $(g,e)$ lies somewhere between $k_1^2$ and $k_2^2$.
We try this kind of numerical quadrature over a wide range of $k^2$.
Since the focus of this study is on the libration of Pluto's $g$, we carry out numerical quadrature just along the positive $y$-axis of the $(e \cos g, e \sin g)$ plane where $g = +\frac{\pi}{2}$.
Note that $\overline{R}(g,e)$'s local extrema can show up at other places than on the $y$-axis in other resonances, in particular in the higher-order resonant systems \citep[e.g.][]{saillenfest2016,saillenfest2017a}.

In the non-resonant case, the numerical averaging can be carried out as a numerical quadrature over the mean longitudes $\lambda$ and $\lambda'$ which are independent and linear functions of time.
However in the case of the resonant objects like Pluto, $\lambda$ and $\lambda'$ are not independent.
In the next section, we briefly summarize how we implement numerical quadrature in the resonant case.

\subsection{Influence of mean motion resonance}\label{ssec:sinusoidapprox}
When a three-body system is librating in a mean motion resonance, the perturbed and perturbing bodies' orbital motions have a recurring pattern with the frequency equal to the encounter frequency.
For Pluto--Neptune's 3:2 exterior mean motion resonance, the encounter period is $T_\mathrm{enc} = \frac{2\pi}{n-n'} \approx 490$ years (approximately three times Neptune's orbital period) where $n'$ is Neptune's mean motion.
The stable resonance center of the critical resonant argument is at $\sigma = 180^\circ$, and $\sigma$ librates around this center with a period $T_\mathrm{lib} \approx 20,000$ years and an amplitude of about 85 to $90^\circ$, as stated in Section \ref{sec:intro}.
Note that the center of the $\sigma$-libration of Pluto remains at $180^\circ$ because Pluto keeps its eccentricity and inclination at moderate values even on gigayear timescales, and its argument of perihelion $g$ librates and does not change much.
We will provide a validation of this statement later in this subsection.
However, for the Plutinos with higher eccentricity or inclination or those that do not exhibit the steady $g$-libration, it is possible that the $\sigma$'s libration center could be displaced from $180^\circ$.

For numerical quadrature of the disturbing function for resonant systems with non-zero libration amplitude of $\sigma$, we recognize two distinct
angles in canonical coordinates: $\lambda'$ and $\sigma$.
Therefore, the double averaging of $R$ is often expressed such as follows \citep[e.g.][their Eq. (12)]{huang2018a}
\begin{equation}
  \overline{R}
  = \frac{1}{4 \pi^2 p} \oint
    \int_0^{2\pi p} R\, d \lambda' d \sigma ,
\label{eqn:def-R-avg-angles}
\end{equation}
where the inner integral is for averaging over $p$ times the period of $\lambda'$, and the outer integral is for averaging over the oscillation period of $\sigma$.
Conceptually, these two angles embody the variations on two fast timescales (fast relative to the secular or vZLK variations of $e$, $I$ and $g$).
Specifically, $\lambda'$ corresponds to $T_\mathrm{enc}$, and $\sigma$ corresponds to its libration period $(T_\mathrm{lib})$ around the resonance center.
$T_\mathrm{lib}$ is significantly longer than $T_\mathrm{enc}$ in general, but both $T_\mathrm{enc}$ and $T_\mathrm{lib}$ are much shorter than the secular timescale that arises for the change of $g$ or $e$ of the perturbed object.
To emphasize the conceptual point regarding the averaging over the two fast timescales, Eq. \eqref{eqn:def-R-avg-angles} can be recast as the double averaging of $R$ over the two time variables as follows:
\begin{equation}
  \overline{R} = \frac{1}{T_\mathrm{lib}} \oint \left( \frac{1}{T_\mathrm{enc}} \oint  R\, d t_\mathrm{enc} \right) d t_\mathrm{lib},
\label{eqn:def-R-avg}
\end{equation}
where the inner integral is for averaging over $T_\mathrm{enc}$, and the outer integral is for averaging over $T_\mathrm{lib}$.
The integration variables for the inner and the outer integral are $t_\mathrm{enc}$ and $t_\mathrm{lib}$, respectively.

We describe below how we numerically implement the integral (quadrature) expressed in Eq. \eqref{eqn:def-R-avg} by making use of the different timescales.
As for the inner integral of Eq.~\eqref{eqn:def-R-avg} (which hereafter is referred to as ${\cal I}_1$), we convert the integration over $t_\mathrm{enc}$ back to the integration over the mean longitude $\lambda'$ of the perturbing body by using the relationship
$ n' d t = d \lambda'$,
together with the range conversion of the variables
(converting $t_\mathrm{enc}$ whose range is from 0 to $T_\mathrm{enc}$
       into $\lambda'$       whose range is from 0 to $2 \pi p$).
Then, we can express ${\cal I}_1$ as follows:
\begin{equation}
\begin{aligned}
{\cal I}_1 \left( g, e, \sigma \right)
 &= \frac{1}{T_\mathrm{enc}} \oint 
    \left. R \right\rvert_{\sigma = \mathrm{constant}}\, dt_\mathrm{enc} \\
 &= \frac{1}{2\pi p} \int_0^{2\pi p}
    \left. R \right\rvert_{\sigma = \mathrm{constant}}\, d\lambda' .
\end{aligned}
\label{eqn:def-I1}
\end{equation}

On the other hand, the outer integral in Eq. \eqref{eqn:def-R-avg} (which hereafter is referred to as ${\cal I}_2$) can be converted into an integral over the sine argument of the critical resonant argument $\sigma$ by making use of
a sinusoid approximation (hereafter referred to as the sinusoid model) as a function of $t_\mathrm{lib}$ as follows:
\begin{equation}
  \sigma(t_\mathrm{lib}) = \sigma_\mathrm{amp} \sin \varepsilon t_\mathrm{lib} + \sigma_0,
\label{eqn:phi-function}
\end{equation}
where
$ \varepsilon = \frac{2\pi}{T_\mathrm{lib}} $
denotes the libration frequency of $\sigma$, $\sigma_\mathrm{amp}$ is the libration amplitude of $\sigma$, and $\sigma_0$ is the value of $\sigma$ at the center of the resonance.
In other words, $\sigma_0$ is the value of $\sigma$ at the exact resonance, and it is $180^\circ$ for Pluto.

The specific value of the frequency $\varepsilon$ does not matter in this study.
We only need to assume that its associated timescale is different: $\varepsilon$ is much smaller than the encounter frequency of Neptune and Pluto $\left( = \frac{2\pi}{T_\mathrm{enc}} \right)$.
Then, using a dimensionless variable $\tau$ defined as
$ \tau = \varepsilon t_\mathrm{lib} $
as the integration variable, we can write the outer integral ${\cal I}_2$ as follows:
\begin{equation}
\begin{aligned}
  {\cal I}_2 \left( g,e \right)
  &= \frac{1}{T_\mathrm{lib}}
     \oint {\cal I}_1  \left( g, e, \sigma \left( t_\mathrm{lib} \right) \right) dt_\mathrm{lib} \\
  &= \frac{1}{2\pi}
     \int_0^{2\pi} {\cal I}_1 
     \left( g, e, \sigma \left( \tau \right) \right) d\tau .
\end{aligned}
\label{eqn:def-I2}
\end{equation}
In Eq. \eqref{eqn:def-I2} we used a relationship
$ \frac{d t_\mathrm{lib}}{d \tau} = \frac{1}{\varepsilon} $,
and another 
 %T_\mathrm{lib} = \frac{2\pi}{\varepsilon}
$ \frac{1}{T_\mathrm{lib}} = \frac{\varepsilon}{2\pi} $,
together with the range conversion of the variables
(converting $t_\mathrm{lib}$ whose range is from 0 to $T_\mathrm{lib}$
       into $\tau$           whose range is from 0 to $\varepsilon T_\mathrm{lib} = 2 \pi$).

Note that, even in the case of
non-zero $\sigma_\mathrm{amp}$ expressed in Eq. \eqref{eqn:phi-function}, we continue to regard the perturbed body's semimajor axis $a$ as a constant in Eq. \eqref{eqn:def-I2}.
In general, the semimajor axis $a$ oscillates along with $\sigma$ except when $\sigma_\mathrm{amp} = 0$
(i.e. except when the object is at the exact resonance).
However, in the case of Pluto, its oscillation amplitude $(\delta a)$ is much smaller than $a$ itself \citep[e.g.,][their Figure 2B in p. 66 is a typical example that shows Pluto's $\delta a/a \ll 1$]{kinoshita1996b}, and in this approximation we neglect its small amplitude.

\begin{figure}[!ht]\centering
 \includegraphics[width=\myfigwidthF]{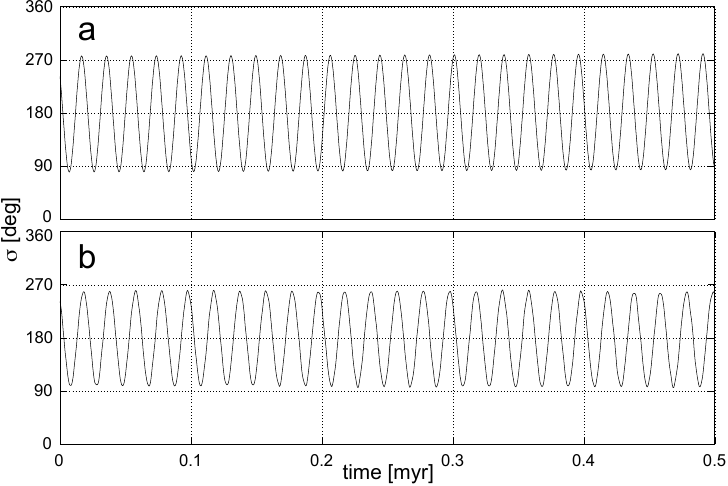}
\caption{%
Time variation of the critical resonant argument of Pluto, $\sigma = 3 \lambda - 2 \lambda' - \varpi$ in Eq. \eqref{eqn:def-sigma}.
\textsf{a}: The Sun--Neptune--Pluto three-body system.
\textsf{b}: The Sun--Jupiter--Saturn--Uranus--Neptune--Pluto system.
\label{fig:figure_np32_sigma}}
\end{figure}

The sinusoidal function form of $\sigma$ in Eq. \eqref{eqn:phi-function} can be validated by direct orbit integration of the equations of motion, although retrofitted.
As an example, we carried out an orbit propagation of Pluto under the influence of Neptune, and another propagation under the influence of all the four giant planets.
For the numerical orbit propagation, we employed the second-order regularized mixed variable symplectic integrator based on the Wisdom--Holman symplectic map \citep{wisdom1991} implemented as a part of the \textsc{SWIFT} package \citep{levison1994}.
We show the time series of $\sigma$ in these two propagations in Fig. \ref{fig:figure_np32_sigma}.
We see that the variation of $\sigma$ can be approximated by a sinusoid with a constant period and a constant amplitude, although the periods and amplitudes of $\sigma$ are slightly different between these two models.
In Appendix \ref{appen:sigma-sinusoid}, we verify that the sinusoid model is a good approximation for Pluto's $\sigma$ not only over a few times the libration period of $\sigma$ but also over secular timescales.
Also in Appendix \ref{appen:sigma-sinusoid},  we clearly see that the location of the resonance center (i.e. the center of the $\sigma$-libration) slightly but certainly varies on the secular timescale.
However, its deviation from $\sigma = 180^\circ$ is modest for Pluto.
This result implies that we can safely neglect the variation of the resonance center in this study.

\begin{figure}[!ht]\centering
 \includegraphics[width=\myfigwidthf]{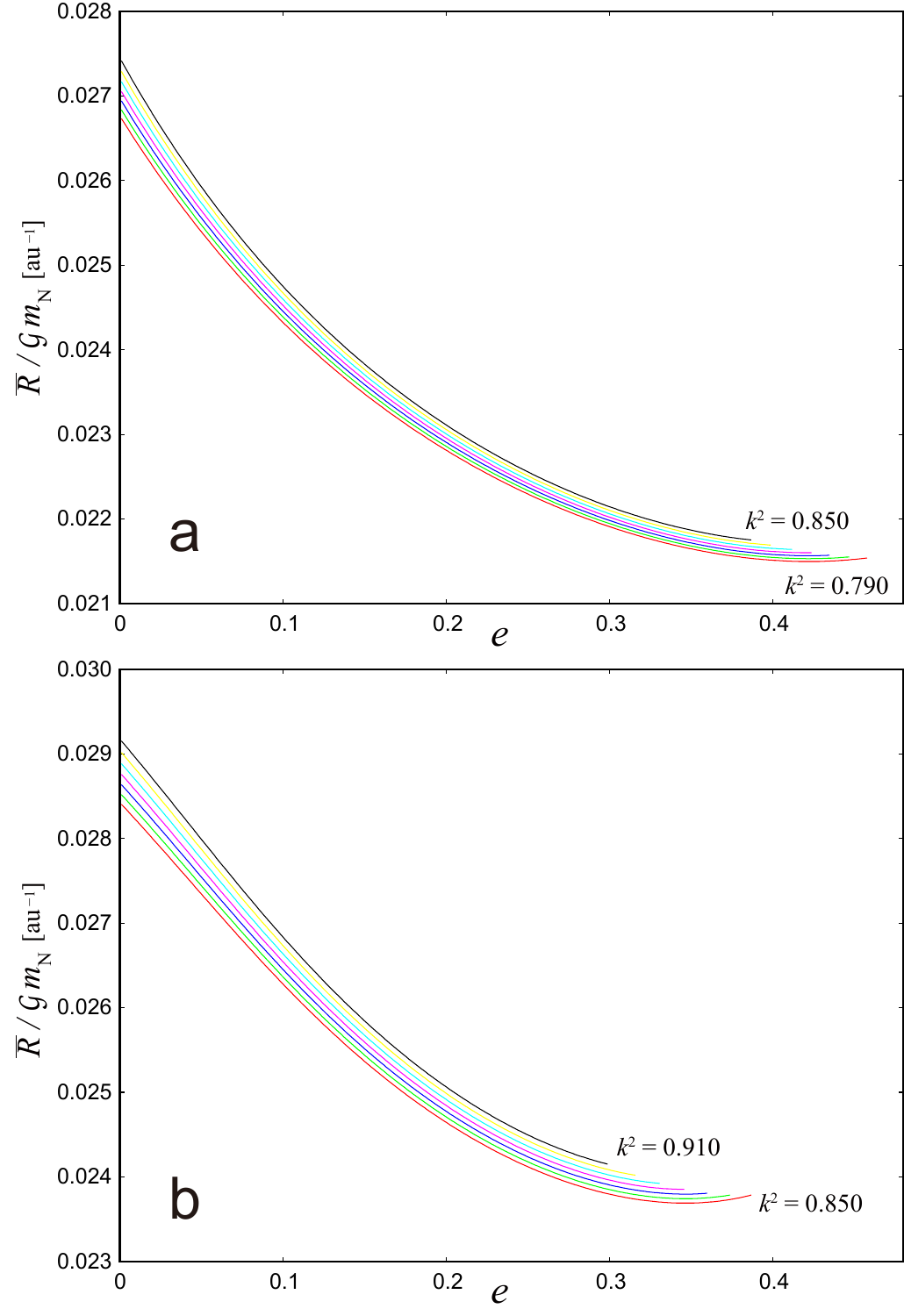}
\caption{%
Values of averaged disturbing function $\overline{R}$ (normalized by ${\cal G} m_\mathrm{N}$) of a Pluto-like object along the direction of $g = \frac{\pi}{2}$ on the $(e\cos g, e\sin g)$ plane as a function of $e$.
$m_\mathrm{N}$ is Neptune's mass.
No other perturbing effect than Neptune is considered.
\textsf{a}: Exact resonant case with $\sigma_\mathrm{amp} = 0$.
Each curve corresponds to the value of $k^2 = 0.790$ (the bottom curve), 0.800, 0.810, 0.820, 0.830, 0.840, and 0.850 (the top curve).
\textsf{b}: When $\sigma_\mathrm{amp} = 90^\circ$.
Each curve corresponds to the value of $k^2 = 0.850$ (the bottom curve), 0.860, 0.870, 0.880, 0.890, 0.900, and 0.910 (the top curve).
}
\label{fig:figure_Ry_J2eq0000}
\end{figure}

In terms of $\sigma$'s amplitude, Pluto's $\sigma_\mathrm{amp}$ is about $85^\circ$ when all the four giant planets are explicitly included in the numerical orbit propagation (Fig. \ref{fig:figure_np32_sigma}b).
This amplitude is known to remain almost the same over the $10^{10}$-year timescale in the actual solar system \citep[][their Figure 14]{ito2002a}.
In our approximate model with the doubly averaged disturbing function, we set $\sigma_\mathrm{amp} = 90^\circ$ for simplicity.
With this constant amplitude, we carry out a sequence of numerical quadrature for the Sun--Neptune--Pluto three-body resonant system using the sinusoid model, and show typical values of the averaged disturbing potential $\overline{R}$ in Fig. \ref{fig:figure_Ry_J2eq0000}.
In this figure, we also show the values of $\overline{R}$ when $\sigma_\mathrm{amp} = 0$ for comparison (the panel \textsf{a}).
In the panel \textsf{a} when $\sigma_\mathrm{amp} = 0$, we find that $\overline{R}$ for $k^2 = 0.790$ has a local minimum.
However, the curve for $k^2 = 0.850$ does not have any local extrema.
Therefore we conclude that $k^2_\mathrm{crit}$ of this system lies between 0.790 and 0.850.
A more detailed series of numerical quadrature using finer steps tells us that $k^2_\mathrm{crit}$ lies between 0.821 and 0.822.
On the other hand, in the panel \textsf{b} when $\sigma_\mathrm{amp} = 90^\circ$, we find that $\overline{R}$ for $k^2 = 0.850$ has a local minimum.
But the curve for $k^2 = 0.910$ does not have any local extrema.
Therefore we conclude that $k^2_\mathrm{crit}$ of this system lies between 0.850 and 0.910.
A more detailed series of numerical quadrature using finer steps tells us that $k^2_\mathrm{crit}$ lies between 0.878 and 0.879.
We will discuss the effect of $\sigma_\mathrm{amp}$ on the value of $k^2_\mathrm{crit}$ in Section \ref{ssec:dependence-sa-J2}.
In Appendix \ref{appen:equiR} we placed pairs of equipotential (equi-$\overline{R}$) contours that correspond to typical cases of Figs. \ref{fig:figure_Ry_J2eq0000}\textsf{a} and \ref{fig:figure_Ry_J2eq0000}\textsf{b} (Figs. \ref{fig:figure_Rxy_j00000_sa000} and \ref{fig:figure_Rxy_j00000_sa090}).
Note that in Fig. \ref{fig:figure_Ry_J2eq0000}, each curve terminates at a different maximum value of the eccentricity.
This is because, for a specified value of $k^2$, there is an upper limit of the perturbed body's eccentricity ($e_\mathrm{max}$) in the averaged CR3BP: $e_\mathrm{max} = \sqrt{1-k^2}$ which takes place when $\cos^2 I = 1$.

When $\sigma_\mathrm{amp}$ is large
(or when $e$ and $I$ are large and vary substantially), i.e. when the object has substantial excursions away from the center of the mean motion resonance, the sinusoid model for $\sigma$ such as in Eq. \eqref{eqn:phi-function} would not be very accurate.
Even the center of the $\sigma$-libration could be substantially displaced from near $180^\circ$ over the timescale of the vZLK oscillation.
In such cases, a more accurate model of $\sigma$ can be made with the use of adiabatic invariants \citep[e.g.][]{sidorenko2014,saillenfest2016,saillenfest2017a,efimov2020a,lei2022,lei2024}.
In the present study, however, our discussion is specific to Pluto.
Pluto's $\sigma_\mathrm{amp}$ is moderate as seen in Fig. \ref{fig:figure_np32_sigma}, and the sinusoid model represents its variation quite well.
Moreover, the corresponding amplitude of its semimajor axis is very small \citep[e.g.][their Figure 2B]{kinoshita1996b}.
This indicates that Pluto does not have substantial excursions away from the center of the 3:2 mean motion resonance with Neptune.
Consequently, the sinusoid model which assumes constant $\sigma_\mathrm{amp}$, provides a good representation.

\section{Influence of other perturbing planets}\label{sec:otherplanets}

In this section, we give considerations on how we can quantify the influence of the other planets on the motion of a Pluto-like perturbed body.

\subsection{The oblate Sun model}\label{ssec:effectiveJ2}
There are several ways to incorporate the gravitational influence of more disturbing planets into the Sun--Neptune--Pluto three-body system.
A simple but efficient way is to express the orbit-averaged perturbation from the planets as a gravitational quadrupole $(J_2)$ added to the point-mass gravity of the Sun.
Specifically, following \cite{malhotra2022}, we add the $J_2$ term to the disturbing function as follows.
Because there are no strong mean motion resonances between Pluto and other planets than Neptune, the secular effects of Jupiter, Saturn, Uranus can be approximated by replacing each planet with a circular ring of radius equal to the perturbing planet's semimajor axis $a'$ and mass equal to the planet's mass $m'$.
This approximation is justified because of the low eccentricity and inclination of these three planets.
The gravitational potential of a ring at a heliocentric distance $r > a'$ and a distance $z$ above the plane of the ring is given by \citep[e.g.][]{jackson1975}
\begin{equation}
  V_\mathrm{ring} = -\frac{G m'}{r} \left[ 1 + \sum_{n=1}^\infty \left(\frac{a'}{r}\right)^{2n} P_{2n}(0)P_{2n} \left( \frac{z}{r}\right) \right], 
\label{e:Vring}
\end{equation}
where $P_{2n}$ is the Legendre polynomial of degree $2n$. 
Here, let us recall that the spatial dependence of the ring potential $V_\mathrm{ring}$ in Eq. \eqref{e:Vring} is similar to the potential exterior to an axially symmetric spheroidal mass, such as that of
a hypothetical oblate model Sun
in the axially symmetric approximation \citep[e.g.][their Eq. (6.218)]{murray1999}
\begin{equation}
  V_{\odot} = -\frac{G m_\odot}{r} \left[ 1 - \sum_{n=1}^\infty J_{2n}\left(\frac{R_\odot}{r}\right)^{2n} P_{2n} \left( \frac{z}{r} \right)\right],
\label{e:Vob}
\end{equation}
where
$R_\odot$ is the Sun's equatorial radius, 
$m_\odot$ is the Sun's mass, and
$J_{2n}$ are the coefficients of the zonal harmonics.
Then, provided that the ring plane (the planetary orbital plane) is identified with the solar equator, comparing Eq. \eqref{e:Vring} and Eq. \eqref{e:Vob} up to $n=1$, we can define the effective $J_2$ of a hypothetical oblate Sun which approximately describes the orbit-averaged potential of a perturbing planet seen by the perturbed object from a distance as follows:
\begin{equation}
  J_\mathrm{2} = \frac{1}{2}\frac{m' a'^2}{m_\odot R_\odot^2} .
\label{e:J2eff}
\end{equation}

The values of $J_\mathrm{2}$ defined in Eq. \eqref{e:J2eff} arising from the inner three giant planets in the actual solar system are given in Table \ref{t:J2eff}.
Using the current masses and time-averaged semimajor axes of Jupiter, Saturn, and Uranus, we find the total effective value for all three planets (Jupiter, Saturn and Uranus), $J_\mathrm{2} \simeq 1574$, as in Table \ref{t:J2eff}.
For obtaining the time-averaged values of $a'$, we consulted the result of the orbit propagation of the four giant planets and Plutinos over 100 myr carried out in \citet{malhotra2025}.
Assuming $J_\mathrm{2} = 1574$, we carry out a similar sequence of numerical quadrature of the resonant three-body system as in the previous section using the disturbing potential, $R - V_{\odot}$.
Note that the conventional sign of $R$ in celestial mechanics is the opposite to the convention in physics \citep[e.g.][]{brouwer1961}, so we have to change the sign of $V_{\odot}$ when we sum them up as $R - V_{\odot}$.

\begin{table}[!ht]
\centering
\caption{
The values of the effective $J_\mathrm{2}$ induced by the inner three giant planets defined as Eq. \eqref{e:J2eff}.
The value of $a'$ in this table are the time-averaged semimajor axis.
For obtaining the values of $J_2$ through Eq. \eqref{e:J2eff}, $R_\odot = 0.004650$ au is adopted from \citet{haberreiter2008}.
}%
\begin{tabular}{cccc}
\toprule
planet & $m_\odot/m'$ & $a'$ (au) & $J_\mathrm{2}$ \\
\midrule
Jupiter  &    1047.3486  &    5.2026  &   597.6  \\
Saturn   &    3497.898   &    9.5551  &   603.6  \\
Uranus   &   22902.98    &   19.2185  &   372.9  \\
Total    &   ---         &   ---      &  1574.1  \\
\bottomrule
\end{tabular}
\label{t:J2eff}
\end{table}

\begin{figure}[!htbp]\centering
 \includegraphics[width=\myfigwidthf]{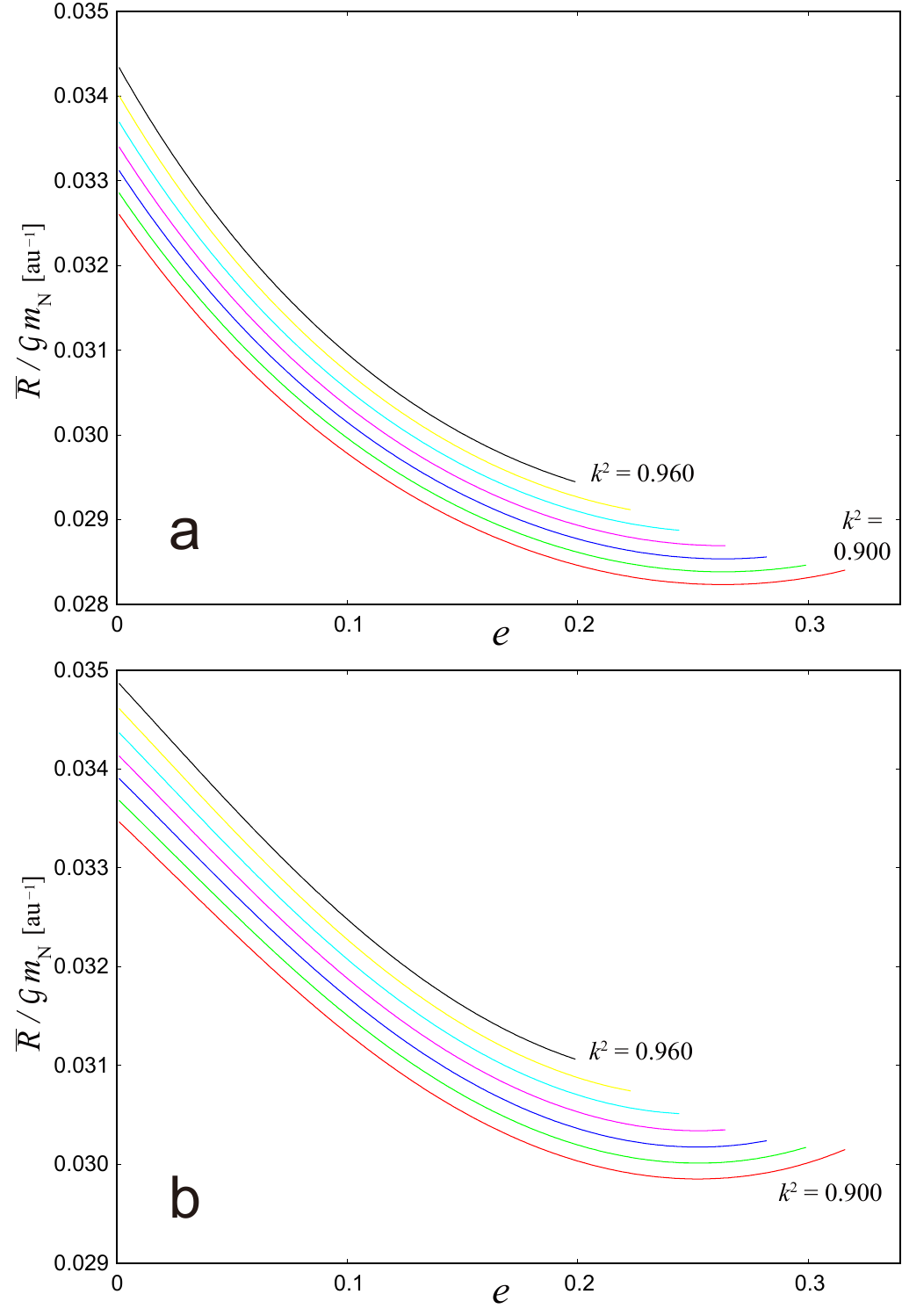}
\caption{%
Same as Fig. \ref{fig:figure_Ry_J2eq0000}, but this figure includes the influence of other planets than Neptune represented by the
oblate model Sun's $J_\mathrm{2} = 1574$.
Meanings of the horizontal and the vertical axes are the same as in Fig. \ref{fig:figure_Ry_J2eq0000}.
\textsf{a}: Exact resonant case with $\sigma_\mathrm{amp} = 0$.
Each curve corresponds to the value of $k^2 = 0.900$ (the bottom curve), 0.910, 0.920, 0.930, 0.940, 0.950, and 0.960 (the top curve).
\textsf{b}: When $\sigma_\mathrm{amp} = 90^\circ$.
Each curve corresponds to the value of $k^2 = 0.900$ (the bottom curve), 0.910, 0.920, 0.930, 0.940, 0.950, and 0.960 (the top curve).
}
\label{fig:figure_Ry_J2eq1600}
\end{figure}

We show typical outcomes of the numerical quadrature in Fig. \ref{fig:figure_Ry_J2eq1600} including a result when $\sigma_\mathrm{amp} = 0$ for comparison (the panel \textsf{a}).
In the panel \textsf{a}, we find that the averaged disturbing potential $\overline{R}$ for $k^2 = 0.900$ has a local minimum, but $\overline{R}$ for $k^2 = 0.960$ does not have any local extrema.
Therefore we conclude that $k^2_\mathrm{crit}$ of this system lies between 0.900 and 0.960.
A more detailed series of numerical quadrature using finer steps tells us that $k^2_\mathrm{crit}$ lies between 0.930 and 0.931.
On the other hand when $\sigma_\mathrm{amp} = 90^\circ$ (the panel \textsf{b}), we find that $k^2_\mathrm{crit}$ lies between 0.935 and 0.936.
Estimated ranges of $k^2_\mathrm{crit}$ is summarized in Table \ref{tbl:k2crit-summary}.
In Appendix \ref{appen:equiR} we placed pairs of equipotential contours that correspond to typical cases of Figs. \ref{fig:figure_Ry_J2eq1600}\textsf{a} and \ref{fig:figure_Ry_J2eq1600}\textsf{b} (Figs. \ref{fig:figure_Rxy_j01574_sa000} and \ref{fig:figure_Rxy_j01574_sa090}).

\begin{table}[!htbp]\centering
\caption{%
Summary of the estimated ranges of $k^2_\mathrm{crit}$ of the Pluto-like object in our study.
The second right column denotes the ranges of the value of $\cos^{-1} \sqrt{k^2_\mathrm{crit}}$ in degrees.
This quantity corresponds to the value of the object's inclination $I$ when $e=0$ at $k^2 = k^2_\mathrm{crit}$.
We cite the corresponding figure numbers in the rightmost column.
``---'' means there is no corresponding figure shown in this article.
}
\begin{tabular}{lccc}
\toprule
\multicolumn{1}{c}{model} & $k^2_\mathrm{crit}$ & $\cos^{-1} \sqrt{k^2_\mathrm{crit}}$ (${}^\circ$) & Fig. \\
\midrule
\multicolumn{4}{c}{$J_\mathrm{2} = 0$} \\
\midrule
resonant $(\sigma_\mathrm{amp} =  0)$        & 0.821--0.822 & 25.03--24.95 & \ref{fig:figure_Ry_J2eq0000}\textsf{a} \\
resonant $(\sigma_\mathrm{amp} = 90^\circ)$  & 0.878--0.879 & 20.44--20.36 & \ref{fig:figure_Ry_J2eq0000}\textsf{b} \\
non-resonant                                 & 0.299--0.300 & 56.85--56.79 & --- \\
\midrule
\multicolumn{4}{c}{$J_\mathrm{2} = 1574$} \\
\midrule
resonant $(\sigma_\mathrm{amp} =  0)$        & 0.930--0.931 & 15.34--15.23 & \ref{fig:figure_Ry_J2eq1600}\textsf{a} \\
resonant $(\sigma_\mathrm{amp} = 90^\circ)$  & 0.935--0.936 & 14.77--14.65 & \ref{fig:figure_Ry_J2eq1600}\textsf{b} \\
non-resonant                                 & 0.219--0.220 & 62.10--62.03 & --- \\
\bottomrule
\end{tabular}
\label{tbl:k2crit-summary}
\end{table}

As a comparison with the resonant cases, we also tested numerical quadrature of the non-resonant, doubly averaged CR3BP by artificially ignoring the mean motion resonance with Neptune but including other perturbing planets represented by
oblate model Sun's $J_\mathrm{2} = 1574$.
The results are added in Table \ref{tbl:k2crit-summary}.
As we see in this table, $k^2_\mathrm{crit}$ of the non-resonant systems is significantly lower than those in the resonant system either when $J_2 = 0$ or $J_2 = 1574$.

\subsection{Dependence of $k^2_\mathrm{crit}$ on $\sigma_\mathrm{amp}$ and $J_2$}\label{ssec:dependence-sa-J2}
\begin{figure}[!htbp]\centering
 \includegraphics[width=\myfigwidthf]{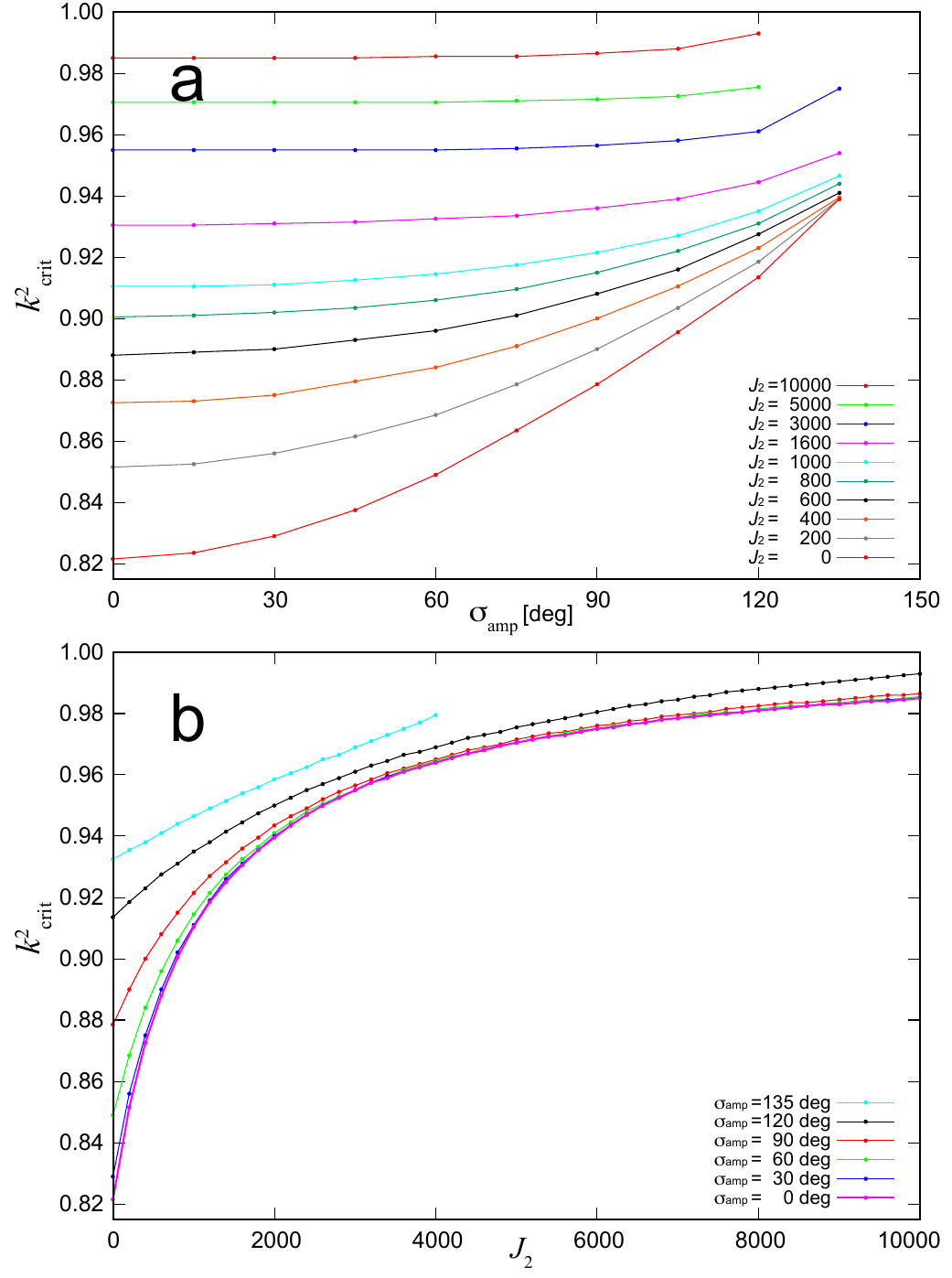}
\caption{%
Dependence of $k^2_\mathrm{crit}$ of the Pluto-like objects on (\textsf{a}) $\sigma_\mathrm{amp}$ and on (\textsf{b}) $J_\mathrm{2}$.
Note that in the panel \textsf{a}, we did not draw all of the curves corresponding to each $J_2$ value shown in the horizontal axis in the panel \textsf{b}.
This is to avoid clutter.
Also in the panel \textsf{b}, we did not plot the curves for $\sigma_\mathrm{amp} = 15^\circ, 45^\circ, 75^\circ, 105^\circ$ to avoid clutter.
}
\label{fig:figure_k2-sa-J2}
\end{figure}

As mentioned in Section \ref{ssec:sinusoidapprox}, choice of $\sigma_\mathrm{amp}$ in the sinusoid model (Eq. \eqref{eqn:phi-function}) affects the value of $k^2_\mathrm{crit}$.
Also, since the oblate model Sun's $J_2$ changes the shape of disturbing potential, $k^2_\mathrm{crit}$ depends on $J_\mathrm{2}$ as well.
By repeating a series of numerical quadrature with various $\sigma_\mathrm{amp}$ and $J_\mathrm{2}$, we obtained Fig. \ref{fig:figure_k2-sa-J2} which illustrates how $k^2_\mathrm{crit}$ of the Pluto-like objects in the 3:2 mean motion resonance with Neptune depends on both the quantities.

Fig. \ref{fig:figure_k2-sa-J2}\textsf{a} shows how $k^2_\mathrm{crit}$ depends on $\sigma_\mathrm{amp}$ at different values of $J_\mathrm{2}$ as a constant parameter.
Note that $J_\mathrm{2} = 0$ means that we do not consider any other disturbing planets than Neptune.
Let us interpret the outcome presented in this panel.
Given a pair of values of $\sigma_\mathrm{amp}$ and $J_\mathrm{2}$, if an object's $k^2$ is located above the corresponding curve in this panel $(k^2 > k^2_\mathrm{crit})$, the object's argument of pericenter $g$ would not show libration.
Here we should recall a fact that $k^2 \leq k^2_\mathrm{crit}$ is a necessary but not sufficient condition for the object's $g$ to librate.
Even when $k^2 \leq k^2_\mathrm{crit}$, the object's $g$ may or may not show libration, depending on the initial value of its disturbing potential.
On the other hand, $k^2 > k^2_\mathrm{crit}$ is a sufficient condition for the object's $g$  to not librate.

To further explain what Fig. \ref{fig:figure_k2-sa-J2}\textsf{a} tells us, let us list three contributions to the time derivative of $g$ (i.e. $\frac{dg}{dt}$) in our model.
The first one arises from the non-resonant, secular terms of the disturbing function.
The second one arises from the resonant terms of the disturbing function.
The third one arises from the contribution of the $J_2$ term to the disturbing function.
Their sum is expressed as follows:
\begin{equation}
  \frac{dg}{dt}
= \left. \frac{dg}{dt} \right\rvert_\mathrm{nr}
+ \left. \frac{dg}{dt} \right\rvert_\mathrm{res}
+ \left. \frac{dg}{dt} \right \rvert_{J_2} ,
  \label{eqn:dgdt-nr-res-J2}
\end{equation}
where the subscripts ``nr'' and ``res'' indicate non-resonant and resonant, respectively.
For small to moderate inclinations, the non-resonant part is positive, i.e. $\left. \frac{dg}{dt} \right\rvert_\mathrm{nr} > 0$.
But it has an inclination dependence such that this rate is smaller when $I$ is larger \citep[e.g.][their Eq. (31) in p. 4858 shows the inclination dependence of $\left. \frac{dg}{dt} \right\rvert_\mathrm{nr}$ as $5\cos^2 I -1$ at the lowest order]{vinson2018}.
Thus, larger $I$ (i.e. small $k^2$) is needed for $\left. \frac{dg}{dt} \right\rvert_\mathrm{nr}$ to diminish.
On the other hand, the resonant part of the disturbing function from the 3:2 mean motion resonance contributes
a rate of $g$ with a dependence as $\left. \frac{dg}{dt} \right\rvert_\mathrm{res} \propto e^{-1}\cos\sigma$ to the lowest order in the perturbed body's eccentricity $e$ \citep[e.g.][their Eq. (8.30)]{murray1999}.
As can be seen from this function form, $\left. \frac{dg}{dt} \right\rvert_\mathrm{res}$ is negative when $\sigma$ librates around $180^\circ$ as in the case of Pluto.
Finally, 
$\left. \frac{dg}{dt} \right \rvert_{J_2}$ is positive unless the inclination of the perturbed body is quite high
\citep[e.g.][their Eq. (11.71) in p. 209 indicates the inclination dependence of $\left. \frac{dg}{dt} \right\rvert_{J_2}$ as $J_2 (5\cos^2 I -1)$ at the lowest order]{prussing2013}.

Having in hand the above properties of each term of $\frac{dg}{dt}$, let us consider the two facts that Fig. \ref{fig:figure_k2-sa-J2}\textsf{a} tells us.
First, we find that $k^2_\mathrm{crit}$ has a trend to monotonically increase with $\sigma_\mathrm{amp}$ at each $J_\mathrm{2}$.
We qualitatively interpret this fact as follows.
\begin{itemize}
\item
As we mentioned above, when $\sigma$ librates around $180^\circ$, the 3:2 mean motion resonance invokes the effect of $\left. \frac{dg}{dt} \right \rvert_\mathrm{res} < 0$.
We should also remember that this term encompasses the factors $\cos \sigma$ and $e^{-1}$.
\item
The factor $\cos \sigma$ embedded in $\left. \frac{dg}{dt} \right \rvert_\mathrm{res}$ takes its most negative value of $-1$ when $\sigma_\mathrm{amp} = 0$ (i.e. when $\sigma$ remains at $\sigma_0 = 180^\circ$).
When $\sigma_\mathrm{amp}$ is larger, the absolute magnitude of $\cos\sigma$ is diminished when averaged over the secular timescales.
In short, the absolute magnitude of $\left. \frac{dg}{dt} \right \rvert_\mathrm{res}$ would be small when $\sigma_\mathrm{amp}$ is large.
On the contrary, the absolute magnitude of $\left. \frac{dg}{dt} \right \rvert_\mathrm{res}$ would be large when $\sigma_\mathrm{amp}$ is small.
\item 
Suppose the value of $\sigma$ is fixed.
Then, the absolute magnitude of $\left. \frac{dg}{dt} \right \rvert_\mathrm{res}$ becomes small when the perturbed object's eccentricity is large.
This is due to the factor $e^{-1}$ that $\left. \frac{dg}{dt} \right \rvert_\mathrm{res}$ contains.
In other words, large eccentricity of the perturbed object contributes to achieve the $g$-libration (i.e. to realize $\frac{dg}{dt} = 0$) by reducing the magnitude of $\left. \frac{dg}{dt} \right \rvert_\mathrm{res}$.
\item 
By the definition of Eq. \eqref{eqn:def-k2}, large eccentricity of the perturbed object leads to $k^2$ being small.
This yields the conclusion that when the magnitude of $\left. \frac{dg}{dt} \right \rvert_\mathrm{res}$ is large due to small $\sigma_\mathrm{amp}$, the value of $k^2$ that achieves the $g$-libration need to be small.
In other words, at small $\sigma_\mathrm{amp}$, $k^2_\mathrm{crit}$ should be small.
Naturally, when $\sigma_\mathrm{amp}$ is large, $k^2_\mathrm{crit}$ is also large.
This trend is typically seen in the curve for $J_2 = 0$ in Fig. \ref{fig:figure_k2-sa-J2}\textsf{a}.
\end{itemize}

Another fact that we find in Fig. \ref{fig:figure_k2-sa-J2}\textsf{a} is that, the dependence of $k^2_\mathrm{crit}$ on $\sigma_\mathrm{amp}$ becomes weaker when $J_\mathrm{2}$ becomes large.
When $J_\mathrm{2}=0$, we see that $k^2_\mathrm{crit}$ increases from about 0.82 to about 0.94 as $\sigma_\mathrm{amp}$ goes from 0 to $135^\circ$.
But when $J_\mathrm{2}$ is larger than 1600, the dependence of $k^2_\mathrm{crit}$ on $\sigma_\mathrm{amp}$ becomes much less clear.
Our interpretation of this fact is as follows.
\begin{itemize}
\item 
When the oblate model Sun's $J_\mathrm{2}$ is large,
$\left.\frac{dg}{dt}\right\rvert_{J_2}$ becomes large, as it grows proportionally with $J_2$ as we mentioned above.
This strongly promotes the precession of the  argument of perihelion in the prograde direction.
\item
In other words, the effect of large $J_2$ (i.e. large $\left. \frac{dg}{dt}\right\rvert_{J_2}$ which is positive) overrides the retrograde motion of $g$ induced by the mean motion resonance (i.e. $\left. \frac{dg}{dt} \right \rvert_\mathrm{res} < 0$) including the effect of $\sigma_\mathrm{amp}$.
\item
Therefore, the dependence of $k^2_\mathrm{crit}$ on $\sigma_\mathrm{amp}$ becomes less prominent when $J_2$ is large.
\end{itemize}

From Fig. \ref{fig:figure_k2-sa-J2}\textsf{b}, we see that when $J_\mathrm{2}$ is small, $k^2_\mathrm{crit}$'s dependence on $\sigma_\mathrm{amp}$ is relatively large (as seen in Fig. \ref{fig:figure_k2-sa-J2}\textsf{a}).
The dependence becomes weakened as $J_\mathrm{2}$ increases.
We can understand this behavior from the facts that we just explained for the panel \textsf{a}.
Note also that the very little difference between the curves for $\sigma_\mathrm{amp} = 0$ and $\sigma_\mathrm{amp} = 30^\circ$ in Fig. \ref{fig:figure_k2-sa-J2}\textsf{b}, and they largely overlap each other over the whole range of $J_2$.
The curves for $\sigma_\mathrm{amp} = 0, 30^\circ, 60^\circ, 90^\circ$ are virtually indistinguishable when $J_2 \gtrsim 3000$.
This indicates that the effect of the other planets (or $J_2$ in this model) largely dominates over the effect of $\sigma_\mathrm{amp}$ in determining the value of $k^2_\mathrm{crit}$ in this system.

Note that although we carried out the series of numerical quadrature by varying the value of $\sigma_\mathrm{amp}$ from 0 to $180^\circ$ in increments of $15^\circ$, we did not observe any local minima along the direction of $g = 90^\circ$ in the disturbing potential when $\sigma_\mathrm{amp} \geq 150^\circ$ (Fig. \ref{fig:figure_k2-sa-J2}\textsf{a}).
In addition, when we set $\sigma_\mathrm{amp} = 135^\circ$, local minima of the disturbing potential show up only when $J_2 \leq 4000$ (Fig. \ref{fig:figure_k2-sa-J2}\textsf{b}).
In other words, it is only in the range of $\sigma_\mathrm{amp} \leq 120^\circ$ that the local minima of the disturbing potential are clearly observed, no matter what the value of $J_2$ is.
We will discuss this point again in Section~\ref{ssec:implication2Pluto} and in Section \ref{sec:discussion}.

%In summary, as seen in Table \ref{tbl:k2crit-summary}, the effect of the 3:2 mean motion resonance is to raise the value of $k^2_\mathrm{crit}$ significantly above that of the non-resonant case, to near Pluto’s current value ($k^2 \approx 0.87$), but still insufficient for Pluto’s $g$ libration. 

\subsection{Implications for Pluto\label{ssec:implication2Pluto}}
In Sections \ref{ssec:dependence-sa-J2} and \ref{ssec:implication2Pluto}, we found that the effect of the additional planets is essential to raise $k^2_\mathrm{crit}$ sufficiently to a value that can support the $g$-libration of the Pluto-like objects in general.
Now let us consider the implications of Fig. \ref{fig:figure_k2-sa-J2} for the $g$-libration of the actual Pluto.
Recall that the actual Pluto in the solar system has $\sigma_\mathrm{amp} \simeq 90^\circ$ and $k^2 \simeq 0.87$.
Examining the red curve for $\sigma_\mathrm{amp} = 90^\circ$ in Fig. \ref{fig:figure_k2-sa-J2}\textsf{b}, we find that the value of $k^2_\mathrm{crit}$ is larger than Pluto's $k^2$ nearly throughout the entire range of $J_2$.
Since $k^2 < k^2_\mathrm{crit}$ seems to almost always hold, one might suspect that Pluto's $g$ would librate no matter how high or low the value of $J_2$ is.
However, this is not true.
As we wrote in Section \ref{ssec:dependence-sa-J2}, the condition $k^2 \leq k^2_\mathrm{crit}$ is a necessary but not sufficient condition for the Pluto-like object's $g$ to librate. 
Depending on which value of disturbing potential the object has (in other words, which equi-$\overline{R}$ contour the object is on), its $g$-libration may take place or may not, even when $k^2 \leq k^2_\mathrm{crit}$.
This is typically illustrated in the right column panels of Fig. \ref{fig:figure_Rxy_j01574_sa090} in Appendix \ref{appen:equiR} where we drew the equi-$\overline{R}$ contours and overlaid Pluto's actual trajectory on those.
From these equi-$\overline{R}$ maps, we see that a Pluto-like object with the same $k^2$ and $\sigma_\mathrm{amp}$ as Pluto would not show the $g$-libration if it has slightly different eccentricity and on a different equi-$\overline{R}$ contour than Pluto.
Therefore, although the Pluto-like objects in the 3:2 mean motion resonance with Neptune that have $k^2 = 0.87$ and $\sigma_\mathrm{amp} = 90^\circ$ could collectively exhibit the $g$-libration for almost any value of $J_2$, the actual Pluto would do it only under certain range of $J_2$ due to its specific value of $\overline{R}$.

\begin{figure}[!ht]\centering
\includegraphics[width=\myfigwidthf]{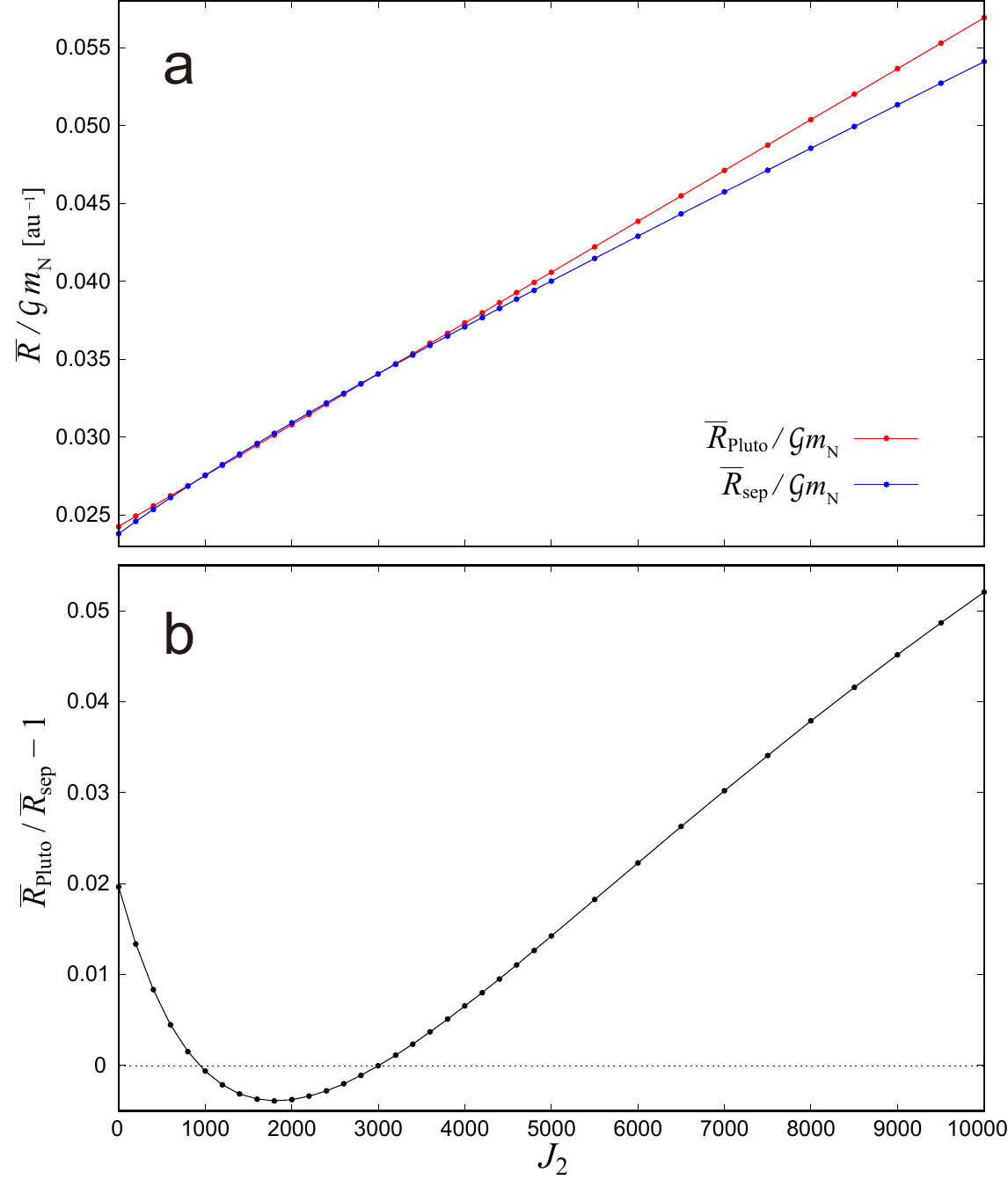}
\caption{%
\textsf{a}:
Dependence of $\overline{R}_\mathrm{Pluto}$ (shown in red) and $\overline{R}_\mathrm{sep}$ (shown in blue) on $J_2$. Both the quantities are normalized by ${\cal G} m_\mathrm{N}$ as in Figs. \ref{fig:figure_Ry_J2eq0000} or \ref{fig:figure_Ry_J2eq1600}.
\textsf{b}:
Fractional difference of the above two quantities defined as $\overline{R}_\mathrm{Pluto} / \overline{R}_\mathrm{sep} - 1$.
Note that some parts of the curve are not smooth and appear bumpy because of the limitation of our approximation method of identifying the location of the separatrix.
}
\label{fig:figure_R_Pluto}
\end{figure}

To quantitatively illustrate the above point, we carry out the following calculation.
\revisedtext{%
For $J_2$ in the range of $0 \leq J_2 \leq 10000$, we first estimate the value of $\overline{R}$ at the location of the actual Pluto in the $(g,e)$ phase space.
We adopt $(g,e) = (113.5^\circ, 0.249)$ for Pluto's approximate location, and refer the value of $\overline{R}$ at this location as $\overline{R}_\mathrm{Pluto}$.
}
In this procedure, we fix the two parameters for Pluto as $k^2 = 0.87$ and $\sigma_\mathrm{amp} = 90^\circ$.
Then, we compare $\overline{R}_\mathrm{Pluto}$ with the value of $\overline{R}$ on the separatrix surrounding the $g$-libration zone in the phase space (i.e. around an equilibrium point).
Hereafter we refer this value as $\overline{R}_\mathrm{sep}$.
While it is straightforward to determine $\overline{R}_\mathrm{pluto}$, estimating the value of $\overline{R}_\mathrm{sep}$ can be challenging, because it is not always easy to find the exact separatrix in the phase space using the output from numerical quadrature.
\marginnote{}
\revisedtext{%
For this purpose we locate the local minimum of $\overline{R}$ along the direction of $g=0$ in the phase space, and consider this minimum value to be $\overline{R}_\mathrm{sep}$.
See Appendix \ref{appen:estimateRsep} for the details of our procedure to estimate $\overline{R}_\mathrm{sep}$.
}

Fig. \ref{fig:figure_R_Pluto}\textsf{a} shows the comparison between $\overline{R}_\mathrm{Pluto}$ (denoted in red) and $\overline{R}_\mathrm{sep}$ (denoted in blue) as a function of $J_2$.
As we observe in this panel, both $\overline{R}_\mathrm{Pluto}$ and $\overline{R}_\mathrm{sep}$ increase with increasing $J_2$.
We also see that $\overline{R}_\mathrm{Pluto}$ is above $\overline{R}_\mathrm{sep}$ when $J_2$ is small and when $J_2$ is large, but $\overline{R}_\mathrm{Pluto}$ is slightly below $\overline{R}_\mathrm{sep}$ in a certain range of $J_2$.
This range is more readily visible when we make a plot of their fractional difference, $\overline{R}_\mathrm{Pluto} /  \overline{R}_\mathrm{sep} - 1$ (Fig. \ref{fig:figure_R_Pluto}\textsf{b}).
Here we find that $\overline{R}_\mathrm{Pluto} < \overline{R}_\mathrm{sep}$ is realized when $1000 \lesssim J_2 \lesssim 3000$.
This is the range of $J_2$ in which the actual Pluto's $\overline{R}$ lies within the $g$-libration zone.

In a previous study, we reported that Pluto's $g$-libration occurs only in the range of $600 \lesssim J_2 \lesssim 3000$ \citep[][their Fig. 5]{malhotra2022}.
\revisedtext{This result} was obtained from a series of numerical orbit propagations of Pluto and Neptune in the solar system including the effect of Sun's $J_2$.
\revisedtext{It} compares favorably with the range of $J_2$ that  we find in Fig. \ref{fig:figure_R_Pluto}\textsf{b}.
\revisedtext{This lends further credence to} the validity of our averaged model for the Pluto--Neptune system.
Moreover, our averaged model elucidates the underlying reasons for the specific range of $J_2$ for the actual Pluto's $g$-libration.

\section{Summary and discussion}\label{sec:discussion}

Our goal in this study is to clarify and quantify the influence of the secular perturbation of the giant planets and the influence of the amplitude of libration of the critical resonant argument on Pluto's $g$-libration.
We confirmed that the additional planets' orbit-averaged effects are well approximated with a quadrupolar potential parameterized with a single parameter, $J_2$, added to the solar gravity, as in our previous work \citep{malhotra2022}. 
Our result shows that, given Pluto's eccentricity and inclination and the libration amplitude of its critical resonant argument, its $g$-libration is only possible because of the combined secular effects of Jupiter, Saturn and Uranus.
In the absence of the secular effects of these three inner giant planets, Pluto's $g$-libration would occur only for much higher values of inclination (or eccentricity) than Pluto's current orbit.

We also found that the $g$-libration of Pluto-like objects is feasible only within the range of $\sigma_\mathrm{amp} \lesssim 120^\circ$ (Fig. \ref{fig:figure_k2-sa-J2}).
When $\sigma_\mathrm{amp} = 135^\circ$, the $g$-libration only occurs when $J_2 \leq 4000$.
When $\sigma_\mathrm{amp} \geq 150^\circ$, we do not observe the $g$-libration at all, no matter what the value of $J_2$ is.
This result is consistent with the numerical results of \citet[][see their Figure 7]{levison1995} which showed that the orbits of Plutinos with $\sigma_\mathrm{amp} \gtrsim 110^\circ$ tend to be unstable.
Their model includes perturbations from the four giant planets, and they carried out a full numerical propagation.
There are other numerical studies that reported the relationship between the value of $\sigma_\mathrm{amp}$ and the orbital stability of the Pluto-like objects which gave similar conclusions as above (e.g., \citet[][their Fig. 5]{duncan1995}, \citet[][their Fig. 5]{nesvorny2000}).
Our result is also consistent with the numerical results of \citet[][see her Figure 6]{malhotra1996} which showed that in Neptune's exterior 3:2 resonance, librating resonant orbits are chaotic when $\sigma_\mathrm{amp} \gtrsim 110^\circ$ when the eccentricity of the perturbed object is $e \gtrsim 0.2$. 
For context, let us note that even in the simple model of the planar CR3BP, the boundary of stable librations of $\sigma$ in Neptune's exterior 3:2 mean motion resonance is quite sensitive to the eccentricity of the Pluto-like massless object \citep{lan2019}.
This boundary is approximately near $120^\circ$ for Pluto-like eccentricities, and it is the boundary above which there is a chaotic zone.
See \citet[][in particular their Figure 11 in which the curve for $e=0.3$ ends near $120^\circ$]{lan2019}.
These results indicate that the stability boundary for Pluto-like orbits lies near $\sigma_\mathrm{amp} \approx 120^\circ$ even without the role of the $g$-libration caused by the vZLK oscillation.
However, it does not necessarily mean that the $g$-libration does not play a significant role in other regions of the phase space neighborhood, such as at higher eccentricities within mean motion resonances.

Our numerical quadrature allows us to explore the effects of $\sigma_\mathrm{amp}$ and of the other planets in some detail.
For example, we now know from Fig. \ref{fig:figure_k2-sa-J2}\textsf{a} that $k^2_\mathrm{crit} \simeq 0.82$ at low values of $\sigma_\mathrm{amp}$ and in the absence of Jupiter, Saturn and Uranus (i.e. $J_2 = 0$).
So, it is possible for Pluto-like resonant TNOs to show the $g$-libration if $I \gtrsim 25^\circ$ (for $e \approx 0$) or $e \gtrsim  0.42$ (for $I \approx 0$). 
Or, with only Jupiter as the additional perturbing planet (i.e. $J_2 \simeq 598$, see Table \ref{t:J2eff}), we have $k^2_\mathrm{crit} \simeq 0.89$ from Fig. \ref{fig:figure_k2-sa-J2}\textsf{a} (see the curve for $J_2 = 600$).
So, the condition for the $g$-libration in this case is $I \gtrsim 20^\circ$ (for $e \approx 0$) or $e \gtrsim 0.34$ (for $I \approx 0$).
This illustrates the sensitivity of the $g$-libration within mean motion resonance to the broader solar system architecture. 

With Sun's $J_2$ as a parametrization for the secular effects of the planets other than Neptune, together with the sinusoid model for the time dependence of $\sigma$, we now have a broad understanding of how $k^2_\mathrm{crit}$ of the Pluto-like objects depends on $J_2$ and on $\sigma_\mathrm{amp}$ (Fig. \ref{fig:figure_k2-sa-J2}).
However, the condition $k^2 \leq k^2_\mathrm{crit}$ is only a necessary but not a sufficient condition for the $g$-libration.
The additional condition is that the value of the averaged disturbing function, $\overline{R}$, must lie within the range of the $g$-libration zone in the phase space $(g,e)$.
We demonstrated that for the actual Pluto’s values of $k^2$ and $\sigma_\mathrm{amp}$, the condition on $\overline{R}$ is satisfied only in the limited range of $1000 \lesssim J_2 \lesssim 3000$ (Fig. \ref{fig:figure_R_Pluto}). 
This range encompasses the value $J_2 = 1574$ representing the approximate orbit-averaged effects of Jupiter, Saturn and Uranus (see Table~\ref{t:J2eff}).
This is how we fully understand the underlying cause of the actual Pluto’s $g$-libration.
% within Neptune’s exterior 3:2 mean motion resonance.

In addition to the 3:2 mean motion resonance with Neptune, other resonances have been noted for Pluto.
One of them is the 1:1 super-resonance in which the difference between longitudes of ascending nodes of Pluto and Neptune oscillates together with the libration of Pluto's $g$ \citep[e.g.][]{williams1971}.
It is possibly one of the so-called secondary resonances \citep[e.g.][]{milani1989a,wan2001}.
Many other secular resonances with timescales in the range of tens of millions of years have been recognized around Pluto's orbit \citep[e.g.][their Fig. 8a]{nesvorny2000}.
In the present work we have not attempted to analyze these other resonances, as our focus is on understanding and quantifying the conditions and circumstances for Pluto's $g$-libration.
We leave to future studies the challenge to identify and quantify the conditions for these weaker resonances working on Pluto and Plutinos.

We end with a few related questions for future investigation:
Can the $g$-libration also occur for objects in non-resonant regions just outside the mean motion resonance region?
And, could objects in such orbits exist in the actual solar system?
If they have planet-crossing eccentricities (like Pluto and many other Plutinos), they would have weaker protection from close encounters and collisions with Neptune.
However, they could be long-lived in high inclination orbits, having the so-called resonance sticking orbits \citep[e.g.][]{tiscareno2009,li2014a}.
This can be tested by telescope observations.
On-going surveys of TNOs will discover more objects in various mean motion resonances with planets, and some of them will have high inclinations \citep[e.g.][]{alexandersen2023,yoshida2024,fraser2024}.

\backmatter

\bmhead{Acknowledgments}
We thank two anonymous reviewers as well as the editor for their detailed and critical feedback.
Their suggestions over many rounds of review significantly improved the presentation and quality of the article.
We are very much in debt to them for their great patience.
Melaine Saillenfest provided us with an intense and fruitful discussion of whether $k^2$ has a lower bound, as well as the usefulness of using adiabatic invariants in this type of study.
TI acknowledges Mami Masuyama for her suggestive discussion of the commonalities between mean motion resonance and the theory of pitch class in music.
Part of the numerical computations presented in this work were carried out at Center for Computational Astrophysics (CfCA), National Astronomical Observatory of Japan.
RM acknowledges funding from NSF (grant AST--1824869), from the program ``Alien Earths'' (supported by the National Aeronautics and Space Administration under Agreement No. 80NSSC21K0593), and from the Marshall Foundation of Tucson, Arizona, USA.
In this study, we used the GNU Scientific Library (GSL) for numerical quadrature which is a collection of routines for high-precision numerical computing.
The authors acknowledge the FreeBSD Project for providing a reliable, open-source operating system that facilitated the computational tasks and data analyses required for this study.
The authors used Overleaf to provide a collaborative and efficient online {\LaTeX} environment, which facilitated the preparation and formatting of this manuscript.
This study has made use of NASA's Astrophysics Data System (ADS) Bibliographic Services.

\section*{Declarations}
The authors declare no competing interests.

\bigskip
\begin{flushleft}%
Editorial Policies for:

\bigskip\noindent
Springer journals and proceedings: \url{https://www.springer.com/gp/editorial-policies}

\bigskip\noindent
Nature Portfolio journals: \url{https://www.nature.com/nature-research/editorial-policies}

\bigskip\noindent
\textit{Scientific Reports}: \url{https://www.nature.com/srep/journal-policies/editorial-policies}

\bigskip\noindent
BMC journals: \url{https://www.biomedcentral.com/getpublished/editorial-policies}
\end{flushleft}

\clearpage

\bibliography{mybib}

\begin{thebibliography}{45}
\providecommand{\natexlab}[1]{#1}
\providecommand{\url}[1]{{#1}}
\providecommand{\urlprefix}{URL }
\providecommand{\doi}[1]{\url{https://doi.org/#1}}
\providecommand{\eprint}[2][]{\url{#2}}
 \bibcommenthead

\bibitem[{{Alexandersen} et~al(2023){Alexandersen}, {Lawler}, {Chen}, {Pike},
  {Kavelaars}, {Peltier}, {Comte}, {Rieger}, {Collyer}, {Morgan}, {Semenchuck},
  and {Holman}}]{alexandersen2023}
{Alexandersen} M, {Lawler} S, {Chen} YT, et~al (2023) {Searching for Pluto's
  neighbors; the large inclination distant objects (LiDO) survey}. In: American
  Astronomical Society Meeting Abstracts, p 104.21,
  \urlprefix\url{https://ui.adsabs.harvard.edu/abs/2023AAS...24110421A}

\bibitem[{Brouwer and Clemence(1961)}]{brouwer1961}
Brouwer D, Clemence GM (1961) Methods of Celestial Mechanics. Academic Press,
  New York, \doi{10.1016/C2013-0-08160-1},
  \urlprefix\url{https://doi.org/10.1016/C2013-0-08160-1}

\bibitem[{Cohen and Hubbard(1965)}]{cohen1965}
Cohen CJ, Hubbard EC (1965) {Libration of the close approaches of Pluto to
  Neptune}. The Astronomical Journal 70:10--13. \doi{10.1086/109674},
  \urlprefix\url{https://doi.org/10.1086/109674}

\bibitem[{{Duncan} et~al(1995){Duncan}, {Levison}, and {Budd}}]{duncan1995}
{Duncan} MJ, {Levison} HF, {Budd} SM (1995) The dynamical structure of the
  kuiper belt. The Astronomical Journal 110:3073--3081. \doi{10.1086/117748},
  \urlprefix\url{https://doi.org/10.1086/117748}

\bibitem[{{Efimov} and {Sidorenko}(2020)}]{efimov2020a}
{Efimov} SS, {Sidorenko} VV (2020) {An analytically treatable model of
  long-term dynamics in a mean motion resonance with coexisting resonant
  modes}. Celestial Mechanics and Dynamical Astronomy 132(5):27.
  \doi{10.1007/s10569-020-09965-5},
  \urlprefix\url{https://doi.org/10.1007/s10569-020-09965-5}

\bibitem[{{Fraser} et~al(2024){Fraser}, {Porter}, {Peltier}, {Kavelaars},
  {Verbiscer}, {Buie}, {Stern}, {Spencer}, {Benecchi}, {Terai}, {Ito},
  {Yoshida}, {Gerdes}, {Napier}, {Lin}, {Gwyn}, {Smotherman}, {Fabbro},
  {Singer}, {Alexander}, {Arimatsu}, {Banks}, {Bray}, {Ramy El-Maarry},
  {Ferrell}, {Fuse}, {Glass}, {Holt}, {Hong}, {Ishimaru}, {Johnson}, {Lauer},
  {Leiva}, {Lykawka}, {Marschall}, {N{\'u}{\~n}ez}, {Postman}, {Quirico},
  {Rhoden}, {Simpson}, {Schenk}, {Skrutskie}, {Steffl}, and
  {Throop}}]{fraser2024}
{Fraser} WC, {Porter} SB, {Peltier} L, et~al (2024) {Candidate distant
  trans-Neptunian objects detected by the New Horizons Subaru TNO survey}. The
  Planetary Science Journal 5(10):227. \doi{10.3847/PSJ/ad6f9e},
  \urlprefix\url{https://doi.org/10.3847/PSJ/ad6f9e},
  {\href{https://arxiv.org/abs/2407.21142}{{https://arxiv.org/abs/arXiv:2407.21142}}}
  {[astro-ph.EP]}

\bibitem[{{Gladman} et~al(2012){Gladman}, {Lawler}, {Petit}, {Kavelaars},
  {Jones}, {Parker}, {Van Laerhoven}, {Nicholson}, {Rousselot}, {Bieryla}, and
  {Ashby}}]{gladman2012}
{Gladman} B, {Lawler} SM, {Petit} JM, et~al (2012) {The resonant
  trans-Neptunian populations}. The Astronomical Journal 144(1):23.
  \doi{10.1088/0004-6256/144/1/23},
  \urlprefix\url{https://doi.org/10.1088/0004-6256/144/1/23},
  {\href{https://arxiv.org/abs/1205.7065}{{https://arxiv.org/abs/arXiv:1205.7065}}}
  {[astro-ph.EP]}

\bibitem[{{Haberreiter} et~al(2008){Haberreiter}, {Schmutz}, and
  {Kosovichev}}]{haberreiter2008}
{Haberreiter} M, {Schmutz} W, {Kosovichev} AG (2008) {Solving the discrepancy
  between the seismic and photospheric solar radius}. {The Astrophysical
  Journal Letters} 675(1):L53. \doi{10.1086/529492},
  \urlprefix\url{https://doi.org/10.1086/529492}

\bibitem[{Hori and Giacaglia(1968)}]{hori1968}
Hori Gi, Giacaglia GEO (1968) Secular perturbations of pluto. In: Giacaglia
  GEO, Oliva WM, Franca LNF, et~al (eds) Research in Celestial Mechanics and
  Differential Equations, CEMC--IMP--USP--67/68--01. Center of Studies in
  Celestial Mechanics, University of Sao Paulo, Sao Paulo, Brazil, pp 4--24

\bibitem[{{Huang} et~al(2018){Huang}, {Li}, {Li}, and {Gong}}]{huang2018a}
{Huang} Y, {Li} M, {Li} J, et~al (2018) {Kozai--Lidov mechanism inside
  retrograde mean motion resonances}. Monthly Notices of the Royal Astronomical
  Society 481(4):5401--5410. \doi{10.1093/mnras/sty2562},
  \urlprefix\url{https://doi.org/10.1093/mnras/sty2562},
  {\href{https://arxiv.org/abs/1809.04959}{{https://arxiv.org/abs/arXiv:1809.04959}}}
  {[astro-ph.EP]}

\bibitem[{Ito and Ohtsuka(2019)}]{ito2019}
Ito T, Ohtsuka K (2019) The {Lidov}--{Kozai} oscillation and {Hugo} von
  {Zeipel}. Monographs on Environment, Earth and Planets 7(1):1--113.
  \doi{10.5047/meep.2019.00701.0001},
  \urlprefix\url{https://ui.adsabs.harvard.edu/abs/2019MEEP....7....1I/abstract},
  {\href{https://arxiv.org/abs/1911.03984}{{https://arxiv.org/abs/arXiv:1911.03984}}}
  {[astro-ph.EP]}

\bibitem[{Ito and Tanikawa(2002)}]{ito2002a}
Ito T, Tanikawa K (2002) Long-term integrations and stability of planetary
  orbits in our solar system. Monthly Notices of the Royal Astronomical Society
  336:483--500. \doi{10.1046/j.1365-8711.2002.05765.x},
  \urlprefix\url{https://doi.org/10.1046/j.1365-8711.2002.05765.x}

\bibitem[{{Jackson}(1975)}]{jackson1975}
{Jackson} JD (1975) {Classical Electrodynamics}. John Wiley {\&} Sons, New
  York, \urlprefix\url{https://www.amazon.co.jp/dp/047143132X}, 2nd edition

\bibitem[{Kinoshita and Nakai(1996{\natexlab{a}})}]{kinoshita1996a}
Kinoshita H, Nakai H (1996{\natexlab{a}}) {Long-term behavior of the motion of
  Pluto over 5.5 billion years}. Earth, Moon, and Planets 72(1--3):165--173.
  \doi{10.1007/BF00117514}, \urlprefix\url{https://doi.org/10.1007/BF00117514}

\bibitem[{Kinoshita and Nakai(1996{\natexlab{b}})}]{kinoshita1996b}
Kinoshita H, Nakai H (1996{\natexlab{b}}) {The motion of Pluto}. In:
  Ferraz-Mello S, Morando B, Arlot JE (eds) Dynamics, Ephemerides, and
  Astrometry of the Solar System. Kluwer Academic Publishers, Dordrecht,
  Boston, London, p 61--70,
  \urlprefix\url{https://ui.adsabs.harvard.edu/abs/1996IAUS..172...61K},
  proceedings of the 172nd Symposium of the International Astronomical Union,
  Paris, France, July 3--8, 1995

\bibitem[{Kozai(1962)}]{kozai1962b}
Kozai Y (1962) Secular perturbations of asteroids with high inclination and
  eccentricity. The Astronomical Journal 67:591--598. \doi{10.1086/108790},
  \urlprefix\url{https://doi.org/10.1086/108790}, note that there is a meeting
  abstract by the same author with the same title published in the same issue
  of the same journal: The Astronomical Journal, vol. 67, 579, 1962,
  \url{https://doi.org/10.1086/108876}

\bibitem[{Lan and Malhotra(2019)}]{lan2019}
Lan L, Malhotra R (2019) {Neptune's resonances in the scattered disk}.
  Celestial Mechanics and Dynamical Astronomy 131(8):39.
  \doi{10.1007/s10569-019-9917-1},
  \urlprefix\url{https://doi.org/10.1007/s10569-019-9917-1},
  {\href{https://arxiv.org/abs/1901.06040}{{https://arxiv.org/abs/arXiv:1901.06040}}}
  {[astro-ph.EP]}

\bibitem[{{Lei}(2024)}]{lei2024}
{Lei} H (2024) {Spin{\textendash}orbit coupling of the primary body in a binary
  asteroid system}. Celestial Mechanics and Dynamical Astronomy 136(5):37.
  \doi{10.1007/s10569-024-10211-5},
  \urlprefix\url{https://doi.org/10.1134/S0038094621050075},
  {\href{https://arxiv.org/abs/2407.21274}{{https://arxiv.org/abs/arXiv:2407.21274}}}
  {[astro-ph.EP]}

\bibitem[{{Lei} et~al(2022){Lei}, {Li}, {Huang}, and {Li}}]{lei2022}
{Lei} H, {Li} J, {Huang} X, et~al (2022) {The von Zeipel--Lidov--Kozai effect
  inside mean motion resonances with applications to trans-Neptunian objects}.
  The Astronomical Journal 164(3):74. \doi{10.3847/1538-3881/ac7c6a},
  \urlprefix\url{https://doi.org/10.3847/1538-3881/ac7c6a},
  {\href{https://arxiv.org/abs/2207.12954}{{https://arxiv.org/abs/arXiv:2207.12954}}}
  {[astro-ph.EP]}

\bibitem[{{Levison} and {Duncan}(1994)}]{levison1994}
{Levison} HF, {Duncan} MJ (1994) The long-term dynamical behavior of
  short-period comets. Icarus 108:18--36. \doi{10.1006/icar.1994.1039},
  \urlprefix\url{https://doi.org/10.1006/icar.1994.1039}

\bibitem[{{Levison} and {Stern}(1995)}]{levison1995}
{Levison} HF, {Stern} SA (1995) {Possible origin and early dynamical evolution
  of the Pluto--Charon Binary.} Icarus 116(2):315--339.
  \doi{10.1006/icar.1995.1128},
  \urlprefix\url{https://doi.org/10.1103/icar.1995.1128}

\bibitem[{Li et~al(2014)Li, Naoz, Kocsis, and Loeb}]{li2014a}
Li G, Naoz S, Kocsis B, et~al (2014) Eccentricity growth and orbit flip in
  near-coplanar hierarchical three-body systems. The Astrophysical Journal
  785:116. \doi{10.1088/0004-637X/785/2/116},
  \urlprefix\url{https://doi.org/10.1088/0004-637X/785/2/116}

\bibitem[{Lidov(1961)}]{lidov1961-en}
Lidov ML (1961) Evolution of the orbits of artificial satellites of planets as
  affected by gravitational perturbation from external bodies. Artificial Earth
  Satellite 8:5--45. Originally in Russian, English translations are available
  as Planetary and Space Science, vol. 9, 719--759, 1962 (DOI
  10.1016/0032-0633(62)90129-0) and AIAA Journal Russian Supplement, vol. 1,
  1985--2002, 1963 (DOI 10.2514/3.54963)

\bibitem[{Malhotra(1996)}]{malhotra1996}
Malhotra R (1996) {The phase space structure near Neptune resonances in the
  Kuiper Belt}. The Astronomical Journal 111:504--516. \doi{10.1086/117802},
  \urlprefix\url{https://doi.org/10.1086/117802},
  {\href{https://arxiv.org/abs/arXiv:astro-ph/9509141}{{https://arxiv.org/abs/arXiv:astro-ph/9509141}}}

\bibitem[{{Malhotra} and {Ito}(2022)}]{malhotra2022}
{Malhotra} R, {Ito} T (2022) {Pluto near the edge of chaos}. Proceedings of the
  National Academy of Science 119(15):2118692119.
  \doi{10.1073/pnas.2118692119},
  \urlprefix\url{https://doi.org/10.1073/pnas.2118692119},
  {\href{https://arxiv.org/abs/2204.04121}{{https://arxiv.org/abs/arXiv:2204.04121}}}
  {[astro-ph.EP]}

\bibitem[{{Malhotra} and {Ito}(2025)}]{malhotra2025}
{Malhotra} R, {Ito} T (2025) {The doubly librating Plutinos}. {The
  Astrophysical Journal} 980(1):115. \doi{10.3847/1538-4357/adacd9},
  \urlprefix\url{https://doi.org/10.3847/1538-4357/adacd9},
  {\href{https://arxiv.org/abs/2501.12345}{{https://arxiv.org/abs/arXiv:2501.12345}}}
  {[astro-ph.EP]}

\bibitem[{{Malhotra} and {Williams}(1997)}]{malhotra1997}
{Malhotra} R, {Williams} JG (1997) {Pluto's heliocentric orbit}. In: {Stern}
  SA, {Tholen} DJ (eds) Pluto and Charon. University of Arizona Press, Tucson,
  pp 127--157,
  \urlprefix\url{https://uapress.arizona.edu/book/pluto-and-charon}

\bibitem[{Milani et~al(1989)Milani, Nobili, and Carpino}]{milani1989a}
Milani A, Nobili AM, Carpino M (1989) {Dynamics of Pluto}. Icarus 82:200--217.
  \doi{10.1016/0019-1035(89)90031-6},
  \urlprefix\url{https://doi.org/10.1016/0019-1035(89)90031-6}

\bibitem[{Murray and Dermott(1999)}]{murray1999}
Murray CD, Dermott SF (1999) Solar System Dynamics. Cambridge University Press,
  Cambridge, \doi{10.1017/CBO9781139174817},
  \urlprefix\url{https://doi.org/10.1017/CBO9781139174817}, known errors are
  published as \url{https://ssdbook.maths.qmul.ac.uk/errors.pdf}

\bibitem[{{Nacozy} and {Diehl}(1978)}]{nacozy1978a}
{Nacozy} PE, {Diehl} RE (1978) {A semianalytical theory for the long-term
  motion of Pluto}. The Astronomical Journal 83:522--530. \doi{10.1086/112231},
  \urlprefix\url{https://doi.org/10.1086/112231}

\bibitem[{{Nesvorn{\'y}} and {Roig}(2000)}]{nesvorny2000}
{Nesvorn{\'y}} D, {Roig} F (2000) {Mean motion resonances in the
  trans-neptunian region. I. The 2:3 resonance with Neptune}. Icarus
  148(1):282--300. \doi{10.1006/icar.2000.6480},
  \urlprefix\url{https://doi.org/10.1006/icar.2000.6480}

\bibitem[{Prussing and Conway(2013)}]{prussing2013}
Prussing JE, Conway BA (2013) {Orbital Mechanics}. Oxford University Press, New
  York,
  \urlprefix\url{https://global.oup.com/ushe/product/orbital-mechanics-9780199837700},
  {Second Edition}

\bibitem[{Saillenfest et~al(2016)Saillenfest, Fouchard, Tommei, and
  Valsecchi}]{saillenfest2016}
Saillenfest M, Fouchard M, Tommei G, et~al (2016) {Long-term dynamics beyond
  Neptune: Secular models to study the regular motions}. Celestial Mechanics
  and Dynamical Astronomy 126(4):369--403. \doi{10.1007/s10569-016-9700-5},
  \urlprefix\url{https://doi.org/10.1007/s10569-016-9700-5},
  {\href{https://arxiv.org/abs/1611.04457}{{https://arxiv.org/abs/arXiv:1611.04457}}}
  {[astro-ph.EP]}

\bibitem[{Saillenfest et~al(2017)Saillenfest, Fouchard, Tommei, and
  Valsecchi}]{saillenfest2017a}
Saillenfest M, Fouchard M, Tommei G, et~al (2017) {Study and application of the
  resonant secular dynamics beyond Neptune}. Celestial Mechanics and Dynamical
  Astronomy 127(4):477--504. \doi{10.1007/s10569-016-9735-7},
  \urlprefix\url{https://doi.org/10.1007/s10569-016-9735-7},
  {\href{https://arxiv.org/abs/1611.04480}{{https://arxiv.org/abs/arXiv:1611.04480}}}
  {[astro-ph.EP]}

\bibitem[{Shevchenko(2017)}]{shevchenko2017}
Shevchenko I (2017) The Lidov--Kozai Effect: Applications in Exoplanet Research
  and Dynamical Astronomy. No. 441 in Astrophysics and Space Science Library,
  Springer International Publishing, Dordrecht,
  \doi{10.1007/978-3-319-43522-0},
  \urlprefix\url{https://doi.org/10.1007/978-3-319-43522-0}

\bibitem[{{Sidorenko} et~al(2014){Sidorenko}, {Neishtadt}, {Artemyev}, and
  {Zelenyi}}]{sidorenko2014}
{Sidorenko} VV, {Neishtadt} AI, {Artemyev} AV, et~al (2014) Quasi-satellite
  orbits in the general context of dynamicsin the 1:1 mean motion resonance:
  perturbative treatment. Celestial Mechanics and Dynamical Astronomy
  120(2):131--162. \doi{10.1007/s10569-014-9565-4},
  \urlprefix\url{https://doi.org/10.1007/s10569-014-9565-4},
  {\href{https://arxiv.org/abs/1409.0417}{{https://arxiv.org/abs/arXiv:1409.0417}}}
  {[astro-ph.EP]}

\bibitem[{{Souami} and {Souchay}(2012)}]{souami2012}
{Souami} D, {Souchay} J (2012) {The solar system's invariable plane}. Astronomy
  and Astrophysics 543:A133. \doi{10.1051/0004-6361/201219011},
  \urlprefix\url{https://doi.org/10.1051/0004-6361/201219011}

\bibitem[{{Tiscareno} and {Malhotra}(2009)}]{tiscareno2009}
{Tiscareno} MS, {Malhotra} R (2009) {Chaotic diffusion of resonant Kuiper Belt
  Objects}. The Astronomical Journal 138:827--837.
  \doi{10.1088/0004-6256/138/3/827},
  \urlprefix\url{https://doi.org/10.1088/0004-6256/138/3/827},
  {\href{https://arxiv.org/abs/0807.2835}{{https://arxiv.org/abs/arXiv:0807.2835}}}

\bibitem[{Vinson and Chiang(2018)}]{vinson2018}
Vinson BR, Chiang E (2018) {Secular dynamics of an exterior test particle: The
  inverse Kozai and other eccentricity--inclination resonances}. Monthly
  Notices of the Royal Astronomical Society 474:4855--4869.
  \doi{10.1093/mnras/stx3091},
  \urlprefix\url{https://doi.org/10.1093/mnras/stx3091}

\bibitem[{Wan et~al(2001)Wan, Huang, and Innanen}]{wan2001}
Wan XS, Huang TY, Innanen KA (2001) {The 1:1 superresonance in Pluto's motion}.
  The Astronomical Journal 121:1155--1162. \doi{10.1086/318733},
  \urlprefix\url{https://doi.org/10.1086/318733}

\bibitem[{Williams and Benson(1971)}]{williams1971}
Williams JG, Benson GS (1971) {Resonances in the Neptune--Pluto system}. The
  Astronomical Journal 76:167--177. \doi{10.1086/111100},
  \urlprefix\url{https://doi.org/10.1086/111100}

\bibitem[{{Wisdom} and {Holman}(1991)}]{wisdom1991}
{Wisdom} J, {Holman} M (1991) {Symplectic maps for the $N$-body problem}. The
  Astronomical Journal 102:1528--1538. \doi{10.1086/115978},
  \urlprefix\url{https://doi.org/10.1086/115978}

\bibitem[{Yoshida et~al(2024)Yoshida, Yanagisawa, Ito, Kurosaki, Yoshikawa,
  Kamiya, Jiang, Stern, Fraser, Benecchi, and Verbiscer}]{yoshida2024}
Yoshida F, Yanagisawa T, Ito T, et~al (2024) {A deep analysis for New Horizons'
  KBO search images}. Publications of the Astronomical Society of Japan
  76(4):psae043. \doi{10.1093/pasj/psae043},
  \urlprefix\url{https://doi.org/10.1093/pasj/psae043},
  {\href{https://arxiv.org/abs/2407.05673}{{https://arxiv.org/abs/arXiv:2407.05673}}}
  {[astro-ph.EP]}

\bibitem[{{Zaveri} and {Malhotra}(2021)}]{zaveri2021}
{Zaveri} N, {Malhotra} R (2021) {Pluto's resonant orbit visualized in 4D}.
  Research Notes of the American Astronomical Society 5(10):235.
  \doi{10.3847/2515-5172/ac3086},
  \urlprefix\url{https://doi.org/10.3847/2515-5172/ac3086}

\bibitem[{von Zeipel(1910)}]{vonzeipel1910}
von Zeipel H (1910) Sur l'application des s\'eries de m. lindstedt \`a
  l'\'etude du mouvement des com\`etes p\'eriodiques. Astronomische Nachrichten
  183(22):345--418. \doi{10.1002/asna.19091832202},
  \urlprefix\url{https://doi.org/10.1002/asna.19091832202}, begr{\"u}ndet von
  H. C. Schumacher, Unter Mitwirkung des Vorstandes der Astronomischen
  Gesellschaft, herausgegeben von Professor Dr. H. Kobold. Band 183, enthaltend
  die Nummern 4369--4392, November 1909 bis M{\"a}rz 1910, Mit drei Tafeln,
  Kiel 1910, Druck von C. Schaidt (Inhaber Georg Oheim, Alfred Oheim). A
  full-text open access PDF file is available from ADS,
  \url{https://ui.adsabs.harvard.edu/abs/1910AN....183..345V}

\end{thebibliography}

\clearpage
\appendix
\section{Behavior of $\sigma$ over secular timescales}\label{appen:sigma-sinusoid}
In Section \ref{ssec:sinusoidapprox} we claimed that the time series of the critical resonant argument $\sigma$ of the 3:2 mean motion resonance between Pluto and Neptune defined in Section \ref{sec:intro} can be well approximated by a sinusoid.
In addition to Fig. \ref{fig:figure_np32_sigma} in Section \ref{ssec:sinusoidapprox}, here we show another figure of the time variation of $\sigma$ over different time periods.

\begin{figure}[!ht]\centering
\includegraphics[width=\myfigwidthF]{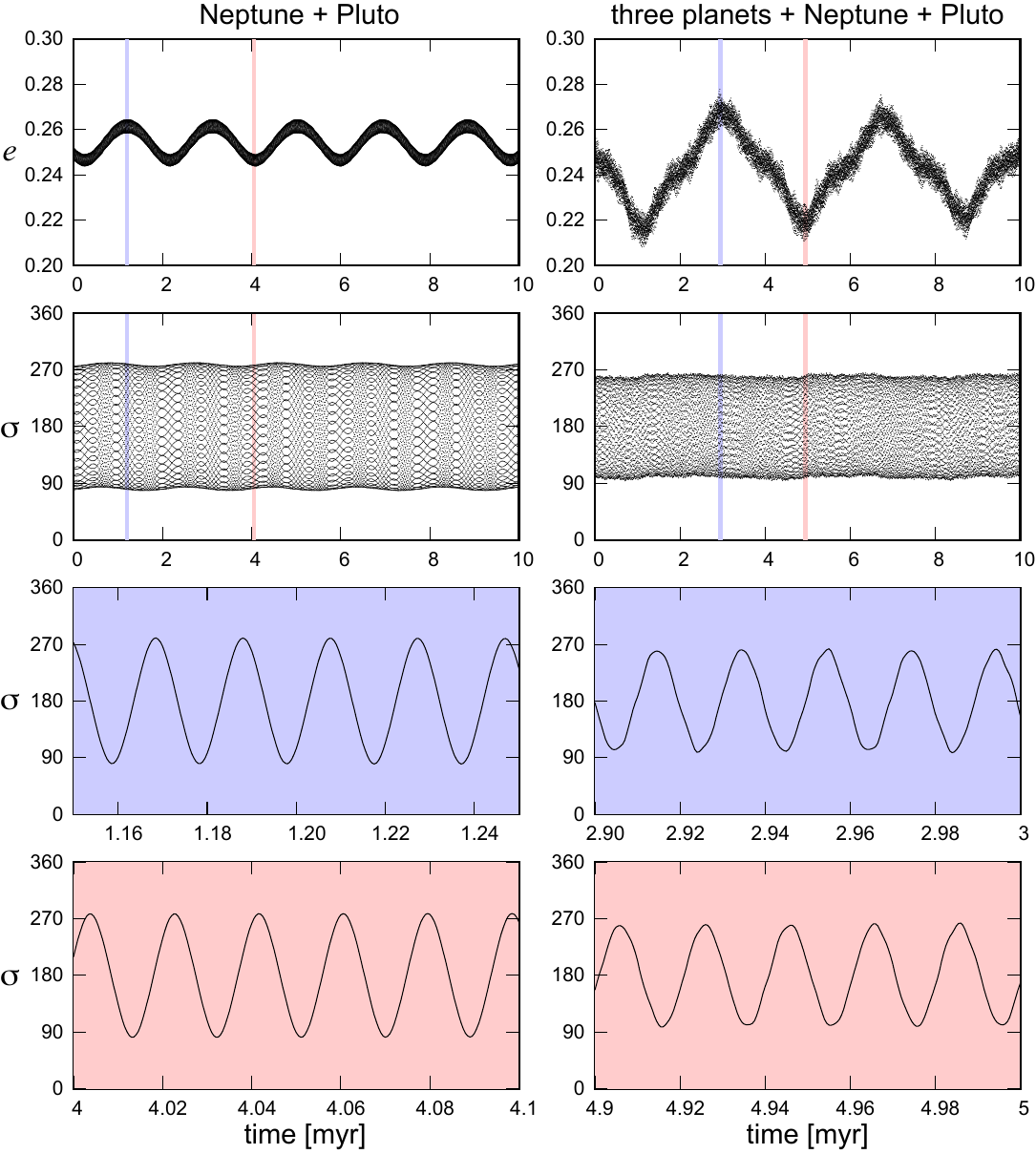}
\caption{%
Time variation of Pluto's eccentricity $(e)$ and $\sigma$ over two different timespan.
The left  column panels are for the Sun--Neptune--Pluto three-body system which corresponds to Fig. \ref{fig:figure_np32_sigma}\textsf{a}.
The right column panels are the Sun--Jupiter--Saturn--Uranus--Neptune--Pluto system which corresponds to Fig. \ref{fig:figure_np32_sigma}\textsf{b}.
(Top) Pluto's $e$, (second top) $\sigma$ in degrees during 10 million years.
In the panels in these two rows, the blue highlighted regions ($t = 1.15$--1.25 myr in the left, and $t=2.90$--3.00 myr in the right) indicate the periods when Pluto's $e$ takes a local maximum.
On the other hand, the red highlighted regions ($t=4.00$--4.10 myr in the left, and $t=4.90$--5.00 myr in the right) indicate those when $e$ takes a local minimum.
(Third top) $\sigma$ during the periods when $e$ takes the local maximum.
(Bottom)    $\sigma$ during the periods when $e$ takes the local minimum.
The third top and the bottom panels are the magnification of the corresponding regions in the second top panels.
}
\label{fig:figure_np32_sigma_more}
\end{figure}

Fig. \ref{fig:figure_np32_sigma_more} shows the time variation of $\sigma$ together with Pluto's eccentricity $e$ over two different timespan.
As in Fig. \ref{fig:figure_np32_sigma}, we employed two systems here:
The left column panels of Fig. \ref{fig:figure_np32_sigma_more} show the results obtained from the Sun--Neptune--Pluto three-body system, and the right column panels show those obtained from the Sun--Jupiter--Saturn--Uranus--Neptune--Pluto system.
In the top and the second top rows, time variation of Pluto's $e$ and $\sigma$ are plotted over the timespan of secular change of orbital elements, i.e. 10 million years.
In the third top and the bottom row panels, they are plotted over the timespan of $T_\mathrm{lib}$, i.e. timescale of the libration of $\sigma$.
The third row panels with the blue background show examples of the variation of $\sigma$ when $e$ takes its local maximum.
The bottom row panels with the red background show those when $e$ takes its local minimum.
The third top and the bottom row panels are the magnification of the corresponding regions in the second top panels.
The result shown in Fig. \ref{fig:figure_np32_sigma_more} strengthens our claim that the variation of $\sigma$ can be approximated by a sinusoid with a constant period and a constant amplitude either we deal with the Sun--Neptune--Pluto three-body system or the 6-body system including other three planets.

Note that \citet[][their Figure 14]{ito2002a} shows the time variation of $\sigma$ over the $\pm 5 \times 10^{10}$ years in their numerical orbit propagation of Pluto and other four giant planets.
In their orbit propagation, Pluto is treated as a planet with mass, not as a massless particle as in this study.

\clearpage
\section{Equipotential contours\label{appen:equiR}}

\begin{figure}[!ht]\centering
\includegraphics[width=\myfigwidthF]{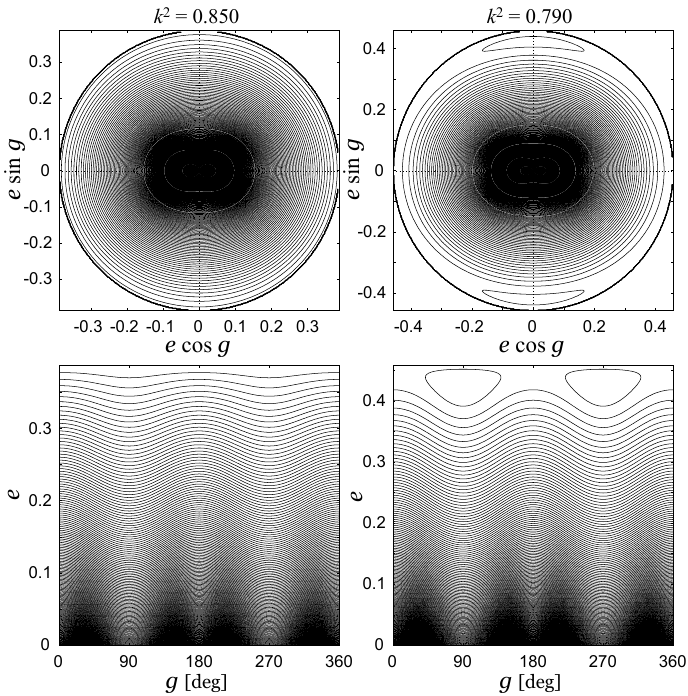}
\caption{%
The equipotential contours for a Pluto-like object in the three-body model.
This figure shows a pair of typical cases in Fig. \ref{fig:figure_Ry_J2eq0000}\textsf{a} where $J_2 = 0$ and $\sigma_\mathrm{amp} = 0$.
(Left  column) when $k^2 = 0.850$,
(right column) when $k^2 = 0.790$.
}
\label{fig:figure_Rxy_j00000_sa000}
\end{figure}

\begin{figure}[!ht]\centering
\includegraphics[width=\myfigwidthF]{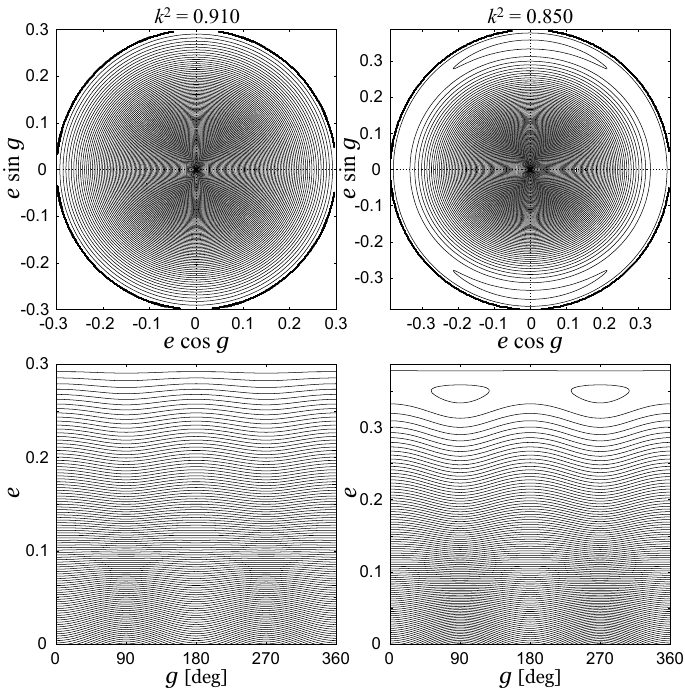}
\caption{%
The same as Fig. \ref{fig:figure_Rxy_j00000_sa000}, but this figure shows a pair of typical cases in Fig. \ref{fig:figure_Ry_J2eq0000}\textsf{b} where $J_2 = 0$ and $\sigma_\mathrm{amp} = 90^\circ$.
(Left  column) when $k^2 = 0.910$,
(right column) when $k^2 = 0.850$.
}
\label{fig:figure_Rxy_j00000_sa090}
\end{figure}

\begin{figure}[!ht]\centering
\includegraphics[width=\myfigwidthF]{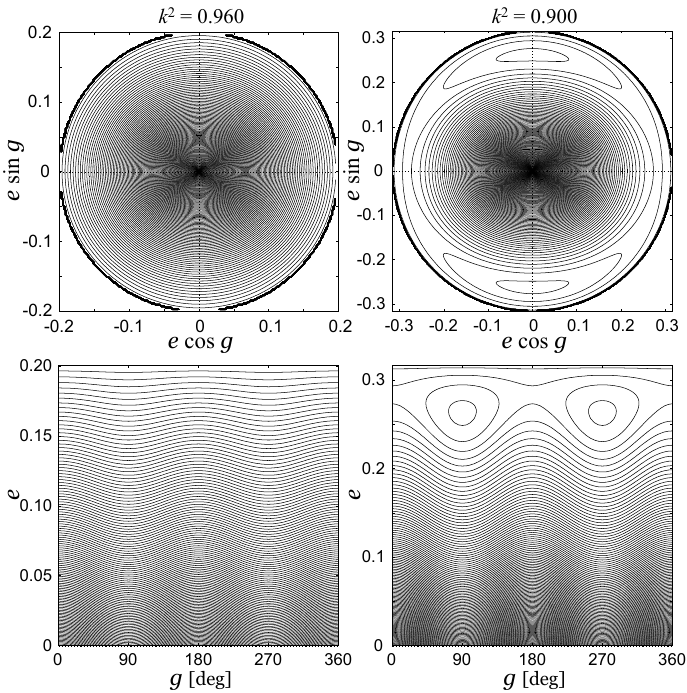}
\caption{%
The same as Figs. \ref{fig:figure_Rxy_j00000_sa000} and \ref{fig:figure_Rxy_j00000_sa090}, but this figure shows a pair of typical cases in Fig. \ref{fig:figure_Ry_J2eq1600}\textsf{a} where $J_2 = 1574$ and $\sigma_\mathrm{amp} = 0$.
(Left  column) when $k^2 = 0.960$,
(right column) when $k^2 = 0.900$.
}
\label{fig:figure_Rxy_j01574_sa000}
\end{figure}

\begin{figure}[!ht]\centering
\includegraphics[width=\myfigwidthF]{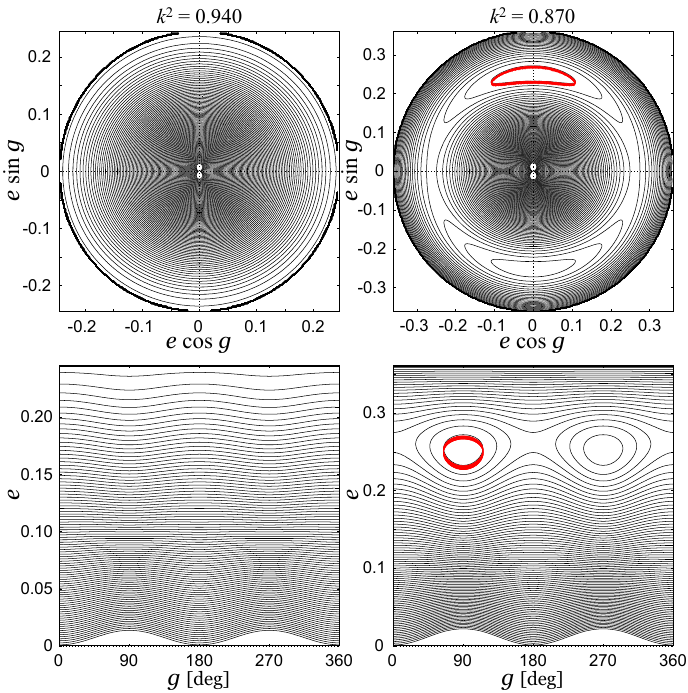}
\caption{%
The same as Figs. \ref{fig:figure_Rxy_j00000_sa000}, \ref{fig:figure_Rxy_j00000_sa090}, and \ref{fig:figure_Rxy_j01574_sa000}, but this figure shows a pair of typical cases in Fig. \ref{fig:figure_Ry_J2eq1600}\textsf{b} where $J_2 = 1574$ and $\sigma_\mathrm{amp} = 90^\circ$.
(Left  column) when $k^2 = 0.940$,
(right column) when $k^2 = 0.870$.
The system shown in the right column panels corresponds to a proxy of the actual Pluto.
What is over-plotted in red in these panels is the result of direct numerical orbit propagation of Pluto under the perturbations from Jupiter, Saturn, Uranus, and Neptune over 20 million years.
Pluto's initial location in the right-column panels is at $(g,e) = (113.5^\circ, 0.249)$.
}
\label{fig:figure_Rxy_j01574_sa090}
\end{figure}

Here we show equipotential contours (equi-$\overline{R}$ curves) of four averaged three-body systems with equilibrium points in $\overline{R}$ and without them (Figs.
\ref{fig:figure_Rxy_j00000_sa000},
\ref{fig:figure_Rxy_j00000_sa090},
\ref{fig:figure_Rxy_j01574_sa000},
\ref{fig:figure_Rxy_j01574_sa090}).
The $k^2$ values employed in these figures correspond to those used in Figs. \ref{fig:figure_Ry_J2eq0000} and \ref{fig:figure_Ry_J2eq1600}.
We plotted the contours on the $(e \cos g, e \sin g)$ plane as well as on the $(g,e)$ plane.

\clearpage
\section{Estimating $\overline{R}_\mathrm{Pluto}$ and $\overline{R}_\mathrm{sep}$\label{appen:estimateRsep}}
Here we explain how we estimate the value of $\overline{R}_\mathrm{Pluto}$ and $\overline{R}_\mathrm{sep}$ that we employed to draw Fig. \ref{fig:figure_R_Pluto} in Section \ref{ssec:implication2Pluto}.
We itemize and describe each step of the procedure.

\begin{figure}[!ht]\centering
\includegraphics[width=\myfigwidthf]{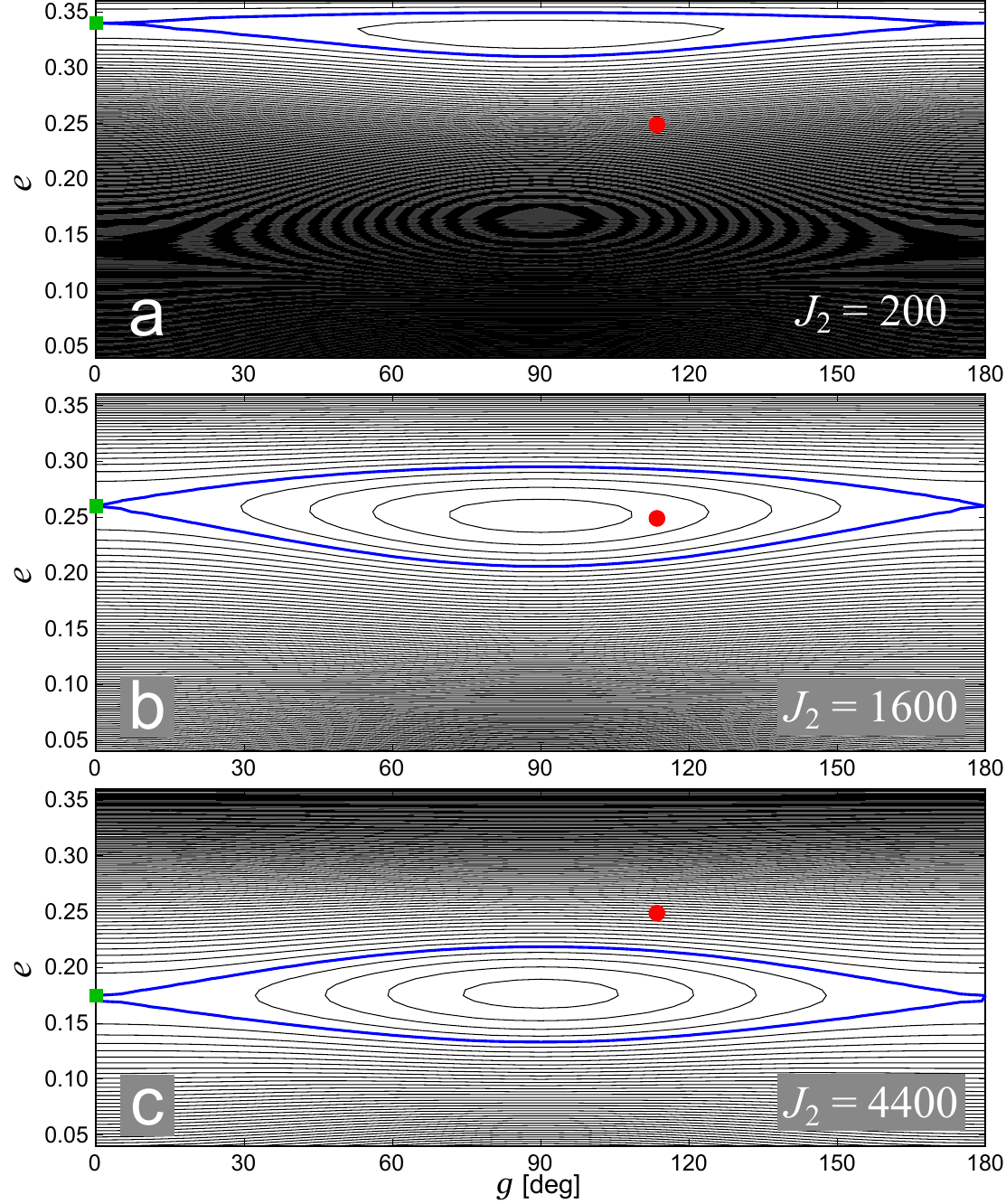}
\caption{%
Equi-$\overline{R}$ contours of the system with $k^2 = 0.87$, $\alpha = 0.763$, and $\sigma_\mathrm{amp} = 90^\circ$ on the $(g,e)$ plane.
\textsf{a}: $J_2 =  200$,
\textsf{b}: $J_2 = 1600$, and
\textsf{c}: $J_2 = 4400$.
$e_\mathrm{max} = \sqrt{1-k^2} \approx 0.36$ in all the three cases.
The red filled circle in each panel denotes the approximate location of the actual Pluto, 
\revisedtext{$(g,e) = (113.5^\circ, 0.249)$.}
We define the value of $\overline{R}$ at this point as $\overline{R}_\mathrm{Pluto}$.
\revisedtext{%
The green filled squares at the left edge in each panel point the locations of local minima of $\overline{R}$ along the direction of $g=0$.
We define the value of $\overline{R}$ at these points as $\overline{R}_\mathrm{sep}$.
We highlight in blue the contours that approximate the separatrix around the equilibrium point in each panel.
These contours converge into the green filled squares when $g=0$.
Note that we do not directly use any of these contours to estimate the values of $\overline{R}_\mathrm{sep}$ or $\overline{R}_\mathrm{Pluto}$.
We draw them to facilitate readers' understanding of the concept that our procedure is based on.
}
}
\label{fig:figure_Rsep_estimate}
\end{figure}

\begin{itemize}
\item
\revisedtext{%
Although the choice of the coordinates does not matter in the following discussion, here we consider the $(g,e)$ plane as the phase space.
It is the same as in the lower panels of Figs \ref{fig:figure_Rxy_j00000_sa000}, \ref{fig:figure_Rxy_j00000_sa090}, \ref{fig:figure_Rxy_j01574_sa000}, and \ref{fig:figure_Rxy_j01574_sa090}.
Using these coordinates, we made Fig. \ref{fig:figure_Rsep_estimate} to facilitate readers' understanding of what we did.
}
\item
\revisedtext{%
Parameters needed for the numerical quadrature to calculate the averaged disturbing potential $\overline{R}$  are $k^2$, $\alpha$, $\sigma_\mathrm{amp}$, and $J_2$.
As for $k^2$ and $\alpha$, we use the current Pluto's values ($k^2 = 0.870$ and $\alpha = 0.763$).
We assume $\sigma_\mathrm{amp} = 90^\circ$ in the sinusoid model as mentioned in previous sections.
As for $J_2$, we use a number of discrete values in the range from $J_2 = 0$ to 10000.}
\item 
First, we consider the point
\revisedtext{$(g,e) = (113.5^\circ, 0.249)$} as the approximate location of Pluto.
In each panel of Fig. \ref{fig:figure_Rsep_estimate}, we marked this location by the red filled circle.
We calculate the value of $\overline{R}$ at this point, and regard it as the proxy of Pluto's $\overline{R}$.
We denote it as $\overline{R}_\mathrm{Pluto}$.
\item 
\revisedtext{%
Next, we carry out a series of numerical quadrature from $e=0$ to $e_\mathrm{max}$ just along the direction of $g=0$.
The purpose of this operation is to locate the local minimum of $\overline{R}$ in this direction.
The green filled squares in Fig. \ref{fig:figure_Rsep_estimate} denote the location of the local minimum in each panel.
We regard these minimum values as $\overline{R}_\mathrm{sep}$.
Note that although we show equi-$\overline{R}$ contours on the $(g,e)$ plane in Fig. \ref{fig:figure_Rsep_estimate}, 
these diagrams are only examples to help the readers comprehend the circumstances in the phase space.
We do not need to perform the numerical quadrature over the entire $(g,e)$ plane just to find the values of $\overline{R}_\mathrm{Pluto}$ or $\overline{R}_\mathrm{sep}$.
\item
The procedure to estimate $\overline{R}_\mathrm{sep}$ mentioned above implicitly assumes that the location of the minimum $\overline{R}$ along the direction of $g = 0$ is a saddle point of the disturbing potential, and the separatrix surrounding the $g$-libration zone emanates from it.
We see the validity of this assumption in Fig. \ref{fig:figure_Rsep_estimate} where the separatrix (the thick blue contour) directly converges into the local minimum of $\overline{R}$ in the direction of $g=0$ (the green filled square).
This type of structure in phase space is visible in the equi-$\overline{R}$ diagrams of some Plutinos in previous studies too (e.g. \citet[][their Figure 11]{gladman2012}, \citet[][their Figure 10(b)]{lei2022}).
}
\item 
Both $\overline{R}_\mathrm{Pluto}$ and $\overline{R}_\mathrm{sep}$ depend on $J_2$.
We estimate their values for each $J_2$ in the range of $0 \leq J_2 \leq 10000$, and plotted them in Fig. \ref{fig:figure_R_Pluto}\textsf{a}.
We showed the fractional difference between $\overline{R}_\mathrm{Pluto}$ and $\overline{R}_\mathrm{sep}$ defined as $\overline{R}_\mathrm{Pluto}/\overline{R}_\mathrm{sep} - 1$ in Fig. \ref{fig:figure_R_Pluto}\textsf{b}.
\end{itemize}

\revisedtext{%
Fig. \ref{fig:figure_Rsep_estimate}\textsf{a} shows an equi-$\overline{R}$ diagram when $J_2 = 200$.
In this case, Pluto's location on this plane is below the lower separatrix and out of the $g$-libration region.
Therefore we have $\overline{R}_\mathrm{Pluto} > \overline{R}_\mathrm{sep}$, and Pluto's $g$ does not librate but circulates.
Fig. \ref{fig:figure_Rsep_estimate}\textsf{b} shows another example when $J_2 = 1600$.
In this case, Pluto's location is between the lower and the upper separatrix, and we have $\overline{R}_\mathrm{Pluto} < \overline{R}_\mathrm{sep}$.
Therefore Pluto's $g$-libration is realized.
Fig. \ref{fig:figure_Rsep_estimate}\textsf{c} is yet another example when $J_2 = 4400$.
In this case, Pluto's location is above the upper separatrix, thus we have $\overline{R}_\mathrm{Pluto} > \overline{R}_\mathrm{sep}$ again and Pluto's $g$ does not librate.
}

Note that, as $J_2$ increases, the location of the equilibrium points of the disturbing potential (with the  same $k^2$, $\alpha$, and $\sigma_\mathrm{amp}$) tends to move toward the region of smaller $e$.
We see this trend in the comparison of the panels \textsf{a}, \textsf{b}, and \textsf{c} of Fig. \ref{fig:figure_Rsep_estimate}.
We can interpret this trend as a compensation of the term of $\left. \frac{dg}{dt} \right\rvert_{J_2}$ which is proportional to $J_2 (5\cos^2 I -1)$ as we mentioned in Section \ref{ssec:dependence-sa-J2}.
When $J_2$ becomes larger, $\cos^2 I$ at the equilibrium point must be smaller to suppress the increase of the absolute value of $\left. \frac{dg}{dt} \right\rvert_{J_2}$ unless $I$ is so large that $\cos^2 I < 1/5$.
Due to the conservation of $k^2 = (1-e^2) \cos^2 I$, the value of $1-e^2$ would become larger.
This means that $e$ at the equilibrium point becomes smaller, which realizes the relocation of the equilibrium point toward the region of smaller eccentricity.

\end{document}